\documentclass[12pt]{iopart}

\pdfoutput=1

\usepackage{tgtermes}

\usepackage{float}
\usepackage{graphicx}  
\usepackage{xcolor}
\usepackage{bm}        
\usepackage{braket}
\usepackage{amssymb}   
\usepackage{epstopdf}
\usepackage{xfrac}
\usepackage{subfigure}
\usepackage{anyfontsize}
\usepackage{array}
\usepackage{multirow}

\usepackage{pifont} 

\usepackage{amsmath}
\usepackage{amsfonts}
\usepackage{bbm}
\definecolor{darkblue}{rgb}{0.1,0.2,0.6}
\definecolor{darkred}{rgb}{0.8,0.1,0.2}
\usepackage[colorlinks,citecolor=darkblue,linkcolor=darkred,urlcolor=darkblue]{hyperref} 
\usepackage[margin=30pt, bf, font=footnotesize, center, justification=justified]{caption}[2004/07/16]
\usepackage{tikz}
\usepackage{enumerate}
\usepackage{setspace}
\usepackage{url}  
\usepackage{mathrsfs}

\usepackage{etoolbox}

\usepackage{lipsum}

\makeatletter
\def\@mkboth#1#2{}
\newlength\appendixwidth
\preto\appendix{\addtocontents{toc}{\protect\patchl@section}}
\newcommand{\patchl@section}{%
  \settowidth{\appendixwidth}{\textbf{Appendix }}%
  \addtolength{\appendixwidth}{1.5em}%
  \patchcmd{\l@section}{1.5em}{\appendixwidth}{}{\ddt}%
}
\makeatother

\def\bZ{\mathbb{Z}}

\catcode`@=11 
\renewcommand\tableofcontents{%
  \section*{\contentsname}%
  \@starttoc{toc}%
}
\catcode`@=12

\usepackage[normalem]{ulem}

\makeatletter
\newcommand{\subscripts}[3]{%
  \@mathmeasure\z@\displaystyle{#2}%
  \global\setbox\@ne\vbox to\ht\z@{}\dp\@ne\dp\z@
  \setbox\tw@\box\@ne
  \@mathmeasure4\displaystyle{\copy\tw@_{#1}}%
  \@mathmeasure6\displaystyle{{#2}_{#3}}%
  \dimen@-\wd6 \advance\dimen@\wd4 \advance\dimen@\wd\z@
  \hbox to\dimen@{}\mathop{\kern-\dimen@\box4\box6}%
}
\makeatother

\begin{document}
\begin{center}{\fontsize{13}{0}
\textbf{
Non-local Order Parameters 
and Quantum Entanglement
for Fermionic Topological Field Theories}}
\end{center}
\begin{center}
Kansei Inamura\textsuperscript{1},
Ryohei Kobayashi\textsuperscript{1,2},
Shinsei Ryu\textsuperscript{3}
\end{center}

{\small
\noindent\textsuperscript{1}  Institute for Solid State Physics, University of Tokyo, Kashiwa, Chiba 277-8583, Japan
\\
\textsuperscript{2}
Center of Mathematical Sciences and Applications, Harvard University, Cambridge, MA 02138
\\
\textsuperscript{3} 
James Franck Institute and Kadanoff Center for Theoretical Physics, University of Chicago, IL~60637
\\}

\begin{center}
\today
\end{center}


\section*{Abstract}
{
We study quantized non-local order parameters,
constructed by using partial time-reversal
and partial reflection,
for fermionic 
topological
phases of matter 
in one spatial dimension 
protected by an orientation reversing symmetry,
using topological quantum field 
theories (TQFTs).
By formulating the order parameters in the Hilbert space of state sum TQFT, we establish 
the connection between the quantized non-local order parameters and the underlying field theory, clarifying the nature of the 
order parameters as topological invariants. We also formulate several entanglement measures including 
the entanglement negativity on state sum spin TQFT, and describe the exact correspondence of the entanglement measures to path integrals on a closed surface equipped with a specific spin structure.
}

\newpage

\vspace{10pt}
\noindent\rule{\textwidth}{1pt}
\tableofcontents
\noindent\rule{\textwidth}{1pt}
\vspace{10pt}

\newpage

\section{Introduction}
\label{sec:intro}

A salient feature of topological phases of matter is the lack of local order parameters characterizing them.
For example, topologically-ordered phases in (2+1)d cannot be characterized by their symmetry-breaking pattern,
but by the anyonic excitations that they support 
\cite{Xiao:803748}.
Quantum Hall systems are characterized by their quantized Hall conductance,
which detects the global topological properties of the ground states. 
Specifically, the Niu-Thouless-Wu formula \cite{Niu-Thouless-Wu1985} 
relates the quantized Hall conductance to the first Chern number defined on a parameter space of boundary conditions, 
through the Berry connection of the (many-body) ground state wave functions in the presence of twisted boundary conditions.

For topological phases of matter beyond the quantum Hall example, 
such as symmetry-protected topological (SPT) phases with internal or spacetime symmetries,
their topological characterization using non-local operations have been proposed
~\cite{Resta1998, Huang2015Detecting, Ryu-Hatsugai2006, PollmannTurner2012, Wen2013invariants, Zaletel2014Detecting}.
For example, in the prototypical case of (1+1)d topological superconductors with
time-reversal symmetry (symmetry class BDI),
the operation called ``partial time-reversal'' (or partial transpose)
\footnote{
  Partial time-reversal and partial transpose may differ by
  a local unitary transformation.
  In this paper, we will exclusively use partial time-reversal.}
can be used to construct a quantized quantity which can detect the $\mathbb{Z}_8$ classification
\cite{FidkowskiKitaev2011, 2010PhRvB..81m4509F}
of (1+1)d BDI topological superconductors
\cite{Shapourian-Shiozaki-Ryu2017a}.
Similarly, one can use ``partial-reflection'' to construct a quantized quantity that can detect
the $\mathbb{Z}_8$ classification of (1+1)d reflection symmetric topological superconductors (symmetry class D$+$R$_-$)
\cite{Shapourian-Shiozaki-Ryu2017a}.
The precise definitions of these quantities will be presented in the later sections.
Henceforth, we loosely call these quantized quantities non-local order parameters.
These are the quantities which are constructed from the ground states of topological phases
by acting with a non-local operation, and detect topological classifications. 
\footnote{
  String-type order parameters, commonly discussed in the context of
  the Haldane phase and related systems,
  are also often called non-local order parameters.
  While string-type order parameters can detect topologically distinct
  phases of matter, they are not quantized.
  In this paper, we will exclusively discuss quantized non-local
  quantities, which are distinct from string-type order parameters.
}
Experimental protocols to measure these quantized non-local order parameters have been proposed
\cite{2019arXiv190605011E}.

The response of quantum Hall systems at low energies and long distances is expected to be described by
the Chern-Simons topological quantum field theory (TQFT).
The Niu-Thouless-Wu formula extracts, from the ground state wave functions, 
the quantized coefficient of the Chern-Simons theory, which is the quantized Hall conductance.
Similarly, (1+1)d time-reversal invariant topological superconductors
are expected to be described by an invertible topological quantum field theory at long
distances~\cite{Kapustin:2014tfa, Freed:2016rqq, Yonekura:2018ufj},
whose partition function on a spacetime gives a bordism invariant.
In this case,
the underlying topological field theory needs to be equipped with a spacetime structure called pin$^-$ structure.
Such invertible pin$^-$ TQFTs are classified up to deformation by the Pontryagin
dual of the pin$^{-}$ bordism group $\Omega_{2}^{\mathrm{pin}^-}(\mathrm{pt})=\mathbb{Z}_8$~\cite{KirbyTaylor}.
It was argued
\cite{Shiozaki-Shapourian-Ryu2017}
that the proposed quantized non-local order parameter
is associated with the partition function of the corresponding invertible pin$^-$ TQFT,
evaluated on a manifold which generates the bordism group
$\Omega_{2}^{\mathrm{pin}^-}(\mathrm{pt})$ (e.g., $\mathbb{RP}^2$),
thereby providing the $\mathbb{Z}_8$-valued topological invariant detecting the classification. 

One of the purposes of this paper is to elucidate the connection between
the quantization of non-local order parameters and the underlying field theory,
clarifying the nature of the non-local order parameters as topological invariants. 
To do this, it is indispensable to formulate
the non-local operations 
in the Hilbert space of TQFT.
This can be achieved
by a local lattice definition of pin$^{-}$ TQFT
recently proposed in~\cite{Gaiotto:2015zta, Kobayashi2019}. 
The lattice
formulation 
makes it possible to construct the ``fixed point" wave functions of fermionic symmetry-protected topological phases on a 1d spatial lattice --
they are the representatives of the ground state wave functions
with shortest possible correlation length,
and have structures akin to Matrix Product States (MPSs).
(See also \cite{Kapustin:2014dxa, Kapustin-Turzillo-You2017,2017PhRvB..95g5108B,Shiozaki-Ryu2017} for relevant references.)
Using the invertible pin$^-$ TQFT generating the $\mathbb{Z}_8$ classification,
we explicitly show that the quantized non-local order parameter
for (1+1)d time-reversal symmetric topological superconductors (class BDI) is
identical to the partition function of the pin$^-$ TQFT computed on $\mathbb{RP}^2$.
Similarly, 
for (1+1)d reflection symmetric topological superconductors (class D$+$R$_-$),
we also prove the exact correspondence between the 
order parameter and the partition function of field theory, based on a lattice definition of pin$^-$ TQFT.

Partial time-reversal (partial transpose) can also be used to construct an entanglement measure for 
mixed quantum states --
the (fermionic) entanglement negativity.
The entanglement negativity 
has been studied recently 
in the context of 
many-body physics 
and quantum field theory
-- see, for example, 
\cite{Calabrese-Cardy-Tonni2012,
Calabrese2013,
Shapourian-Shiozaki-Ryu2017b,
2019PhRvA..99b2310S,
2019JSMTE..04.3106S,
PhysRevA.88.042319,
PhysRevA.88.042318,
Wen2016_2,Wen2016_1}.
Experimental protocols for the entanglement negativity
has been also proposed \cite{2019PhRvA..99f2309C}.
The formalism which we will develop in this paper allows us to study the entanglement negativity in 2d fermionic TQFT,
i.e., in the fixed point
wave functions of fermionic symmetry protected topological phases.
While not expected to be a topological invariant,
the entanglement negativity in the fixed point wave functions
are known to take specific values.
For example, for the ground state of the (1+1)d Kitaev chain in its topologically 
non-trivial phase, 
the entanglement negativity for adjacent intervals 
is given by $\log \sqrt{2}$, 
which is related to the quantum dimension
of the boundary Majorana modes.
We will reproduce this result by using our  TQFT/MPS formalism.

Another quantity of our interest is
the moments of the partially-transposed  reduced density matrix,
which give the spectrum of 
partially-transposed reduced density matrix
(``the negativity spectrum") 
\cite{Ruggiero_2016, Mbeng_2017, Shapourian-Ruggiero-Ryu-Calabrese2019}.
Just like the entanglement spectrum provides 
the universal information of 
quantum ground states (for gapped quantum states in particular),
we expect that we could extract 
universal (topological) data from the negativity spectrum. 
For comparison, it is good to recall that 
for the case of unitary and on-site symmetry, 
ground states of symmetry-protected phases in 
one spatial dimension are characterized by 
symmetry-protected degeneracy of their entanglement spectra 
\cite{Ryu-Hatsugai2006,Li2008,Pollmann-Berg-Turner-Oshikawa2010,Turner2010}.
Here, we will show that we can develop a similar 
diagnostics by using the negativity spectrum
for fermionic symmetry-protected topological phases
protected by time-reversal.

\subsection{Summary of results}
The first column of Table~\ref{tab:list} lists the quantities studied in this paper.
We will prove that these quantities
computed for the fixed point wave function
of (1+1)d topological superconductors (constructed from lattice TQFT)
exactly give the partition functions
of the TQFT on spacetime manifolds
listed in the second column.
The third column lists the explicit values of these quantities.

In the first and second rows of Table~\ref{tab:list},
we find that the partial time-reversal and partial reflection on 
the fixed-point wave function
gives the partition function of pin$^-$ TQFT on $\mathbb{RP}^2$.
Here, although the state is initially prepared on a boundary of an oriented surface (e.g., disk),
we will see that the process of partial time-reversal or reflection introduces
a combinatorial pin$^-$ structure in the whole triangulated spacetime,
which becomes unoriented,
providing a pin$^-$ bordism invariant.
Especially, when computed within
the lattice TQFT generating the $\mathbb{Z}_8$ classification, the bordism invariant corresponds to the Arf-Brown-Kervaire invariant.

In the third 
row
of Table~\ref{tab:list},
we find that the exponential of the R\'enyi entanglement negativity
$e^{\mathcal{E}_{n}}$ of the fixed point wave function
is equal to  
the TQFT partition function on a closed oriented manifold with genus $(n/2-1)$.
The notation like $(n/2-1)\times$ (NS, NS) means that the spacetime manifold has
an induced spin structure
given by a connected sum of $(n/2-1)$ copies of (NS,NS) torus.
When evaluated for
the correctly-normalized fixed point wave function, 
the TQFT partition function
has the Euler term, which makes the R\'enyi negativity
proportional to the Euler characteristic
of the spacetime manifold up to constant.

In the fourth 
row of Table~\ref{tab:list},
we present the moment $Z_n$ of a ground state density matrix acted by partial
time-reversal.
We find an interesting periodicity of the spacetime structure with respect to the degree of the moment.
The spin structure of the spacetime manifold is fixed for each $n$, and it has a
pattern of mod 4 periodicity.
Entries marked as ``-'' in Table~\ref{tab:list} mean that the induced structure on a spacetime manifold is not spin.
We will see that these cases have a vortex of fermion parity introduced in a spacetime, which makes a spin structure ill-defined.
When evaluated
for the fixed point wave function, 
the phase of the TQFT partition function corresponds to the Arf invariant on a
spacetime manifold equipped with a spin structure,
which has a pattern of mod 8 periodicity.

Although most of the analysis will be done for a specific pin$^-$ invertible
TQFT in the main text which corresponds to the Kitaev chain,
the partial time-reversal and reflection can also be formulated on the Hilbert space of generic spin/pin TQFT prepared by $\mathbb{Z}_2$-graded Frobenius algebra. The result presented in this paper can be safely generalized for generic spin/pin TQFT on a lattice.

The rest of the paper is organized as follows.
In Sec.~\ref{sec:tqft}, we review the lattice construction of fermionic TQFT on
unoriented spacetime manifolds.
In Sec.~\ref{sec:pt}, we formulate partial time-reversal for the Hilbert space
of pin$^-$ TQFT.
In Sec.~\ref{sec:nega}, we discuss the formulation of the entanglement
negativity.
In Sec.~\ref{sec:moment}, we discuss the formulation of 
the moments of 
the density matrices with
partial time-reversal and the periodicity presented in Table~\ref{tab:list}.
Finally, in Sec.~\ref{sec:pr}, we illustrate the partial reflection.

\begin{table}[!]
\begin{center}
\caption{
\label{tab:list}
List of many-body order parameters for fermion SPT phases formulated 
in lattice TQFT. 
}
\begin{tabular}{| >{\centering\arraybackslash}m{3.4cm}  |
>{\centering\arraybackslash}m{1.1cm}|
>{\centering\arraybackslash}m{4cm} | >{\centering\arraybackslash}m{3cm} | >{\centering\arraybackslash}m{1cm} | }
\hline
\multicolumn{2}{|c|}{Order parameter} &  Spacetime manifold & the Kitaev chain  & Sec. \\
\hline
\multicolumn{2}{|c|}{partial time-reversal }
&  \multirow{2}{*}{$\mathbb{RP}^2$} & \multirow{2}{*}{$\frac{1}{2\sqrt{2}} e^{\pm 2\pi i / 8}$} & \multirow{2}{*}{\ref{sec:pt}} \\
\multicolumn{2}{|c|}{$\mathrm{tr}_{I}(\rho_I\rho_{I}^{\mathcal{T}_1})$}
&  \ & \ & \ \\

\hline
\multicolumn{2}{|c|}{partial reflection}
&  \multirow{2}{*}{$\mathbb{RP}^2$} & \multirow{2}{*}{$\frac{1}{\sqrt{2}} e^{\pm 2\pi i / 8}$} & \multirow{2}{*}{\ref{sec:pr}} \\
\multicolumn{2}{|c|}{$\bra{\psi}\mathcal{R}_{\mathrm{part}}\ket{\psi}$}
&  \ & \ & \ \\

\hline
\multicolumn{2}{|c|}{R\'enyi entanglement}  & \ & \ & \ \\
\multicolumn{2}{|c|}{negativity}  & $(n/2-1)\times$ (NS, NS)  &  $(2\sqrt{2})^{2-n}\times \frac{1}{2}$ & \ref{sec:nega} \\
\multicolumn{2}{|c|}{$e^{\mathcal{E}_{n}}$}
&  \ & \ 
& \ \\

\hline
\ & $n=0$ mod $8$ &  $(n/4)\times$ (R, R) $+$ $(n/4-1)\times$ (NS, R)  & $(2\sqrt{2})^{2-n}\times\frac{1}{2}$ & \ \\
\cline{2-4}
\ & $n=1$ mod $8$ &  $(n-1)/4\times$ (R, R) $+$ $(n-1)/4\times$ (NS, R)  &$(2\sqrt{2})^{1-n}\times 1$ & \ \\
\cline{2-4}
\ & $n=2$ mod $8$ &  -  & 0 & \ \\
\cline{2-4}
moment of partial time-reversal & $n=3$ mod $8$ &  $(n+1)/4\times$ (R, R) $+$ $(n-3)/4\times$ (NS, R)  & $(2\sqrt{2})^{1-n}\times (-1)$ & \multirow{2}{*}{\ref{sec:moment}} \\
\cline{2-4}
$Z_n$ & $n=4$ mod $8$ &  $(n/4)\times$ (R, R) $+$ $(n/4-1)\times$ (NS, R)  &$(2\sqrt{2})^{2-n}\times \left(-\frac{1}{2}\right)$ & \ \\
\cline{2-4}
\ & $n=5$ mod $8$ &  $(n-1)/4\times$ (R, R) $+$ $(n-1)/4\times$ (NS, R)  &$(2\sqrt{2})^{1-n}\times (-1)$ & \ \\
\cline{2-4}
\ & $n=6$ mod $8$ &  -  & 0 & \ \\
\cline{2-4}
\ & $n=7$ mod $8$ &  $(n+1)/4\times$ (R, R) $+$ $(n-3)/4\times$ (NS, R)  & $(2\sqrt{2})^{1-n}\times 1$ & \ \\
\hline
moment of partial time-reversal & $n=0$ mod $2$ &  $(n/2-1)\times$ (NS, NS)  & $(2\sqrt{2})^{2-n}\times \frac{1}{2}$ & \multirow{2}{*}{\ref{sec:moment}} \\
\cline{2-4}
$\widetilde{Z}_n$ (twisted) & $n=1$ mod $2$ & - &0 & \ \\
\hline
\end{tabular}
\end{center}
\end{table}

\section{Review of fermionic TQFT}
\label{sec:tqft}

In this section, we recall the lattice construction of the spin and pin$^\pm$ TQFT on a 2d manifold $M$,  
following~\cite{Gaiotto:2015zta, Kobayashi2019}. 
We provide a recipe to construct a state sum definition of spin/pin TQFT, by formulating the spin/pin theory called 
the Gu-Wen Grassmann integral on $M$, equipped with a $\bZ_2$ global symmetry, whose partition function has the form~\cite{Gaiotto:2015zta, Gu:2012ib, KOT2019}
\begin{equation}
z[M, \eta,\alpha]=\sigma(M, \alpha)(-1)^{\int_M\eta\cup\alpha},
\label{eq:pintft}
\end{equation}
where $\alpha\in Z^{1}(M, \bZ_2)$ is a background $\mathbb{Z}_2$
gauge field of the $\mathbb{Z}_2$ symmetry, and $\eta$ specifies a spin or pin$^{\pm}$ structure on $M$, which is related to the obstruction of the structure as $\delta\eta=w_2$ (resp.~$\delta\eta=w_2+w_1^2$) in the spin or pin$^+$ (resp.~pin$^-$) case.
Here, $w_{1,2}$ are the first an second 
Stiefel-Whitney classes, respectively.

$\sigma(M, \alpha)$ is written in terms of a certain path integral of Grassmann variables defined by giving a triangulation of $M$.
(In the following, when there is no confusion, 
we simply write $z[\eta, \alpha]$, $\sigma(\alpha)$, instead of 
$z[M,\eta,\alpha]$, $\sigma(M,\alpha)$, etc.)

By studying the effect of re-triangulations and gauge transformations, this theory is shown to be anomaly free for a spin or pin$^{-}$ surface which we focus on in the main text of the present paper.
Then, one can construct a spin or pin$^{-}$ theory fully invariant under the change of triangulation and gauge transformations,
by coupling the Grassmann integral with an anomaly free bosonic theory
$\widetilde{Z}_M[\alpha]$ called a ``shadow theory''~\cite{Thorngren2018bosonization, Bhardwaj2017Statesum, Ellison2019}, and then gauging the $\mathbb{Z}_2$ symmetry,
\begin{equation}
Z[M, \eta]=\sum_{\alpha}z[M, \eta, \alpha]\widetilde{Z}_M[\alpha].
\label{eq:gaugedpin}
\end{equation}

The rest of this section is organized as follows. In Sec.~\ref{subsec:spintft} and \ref{subsec:pintft}, we review the construction of the Grassmann integral on a spin and pin surface respectively. In Sec.~\ref{subsec:ABK}, we provide the lattice construction of a pin$^-$ invertible TQFT, which describes
(1+1)d topological superconductors in class BDI
at long distances.
Then, we describe the construction of spin/pin TQFT 
on a surface with a non-empty boundary in Sec.~\ref{subsec:wfn},
and apply it to construct 
the "fixed-point"  ground state wave function of 
(1+1)d topological superconductors
in Sec.~\ref{subsec:kitaev}.
This is the ground state wave function 
of the Kitaev chain 
deep inside its topological superconductor phase,
with the smallest correlation length.

\subsection{spin TQFT on the lattice}
\label{subsec:spintft}
We endow an oriented surface $M$ with a triangulation. In addition, we take the barycentric subdivision for the triangulation of $M$. Namely, each 2-simplex in the initial triangulation of $M$ is subdivided into 6 simplices, whose vertices are barycenters of the subsets of vertices in the 2-simplex. We further assign a local ordering to vertices of the barycentric subdivision, such that a vertex on the barycenter of $i$ vertices is labeled as $i$.

Each simplex can then be either a $+$ simplex or a $-$ simplex, depending on whether the ordering agrees with the local orientation or not.
We assign a pair of Grassmann variables 
$\theta_e, \overline{\theta}_e$ on each 1-simplex $e$ of $M$ when
$\alpha(e)=1$, we associate $\theta_e$ on one side of $e$ contained in one of 2-simplices neighboring $e$ (which will be specified later), and $\overline{\theta}_e$ on the other side. Then, $\sigma(M, \alpha)$ is defined as
\begin{equation}
    \sigma(M, \alpha)=\int\prod_{e|\alpha(e)=1}d\theta_e d\overline{\theta}_e \prod_t u(t),
    \label{sigmadefspin}
\end{equation}
where $t$ denotes a 2-simplex, and $u(t)$ is the product of Grassmann variables contained in $t$.
Namely, $u(t)$ on $t=(012)$ is the product of
$\vartheta_{12}^{\alpha(12)}, \vartheta_{01}^{\alpha(01)}, \vartheta_{02}^{\alpha(02)}$. 
Here, $\vartheta$ denotes $\theta$ or $\overline{\theta}$ depending on the choice of the assigning rule, which will be discussed later. The order of Grassmann variables in $u(t)$ will also be defined shortly.
We note that $u(t)$ is ensured to be Grassmann-even when $\alpha$ is closed.

\begin{figure}[htb]
\centering
\includegraphics{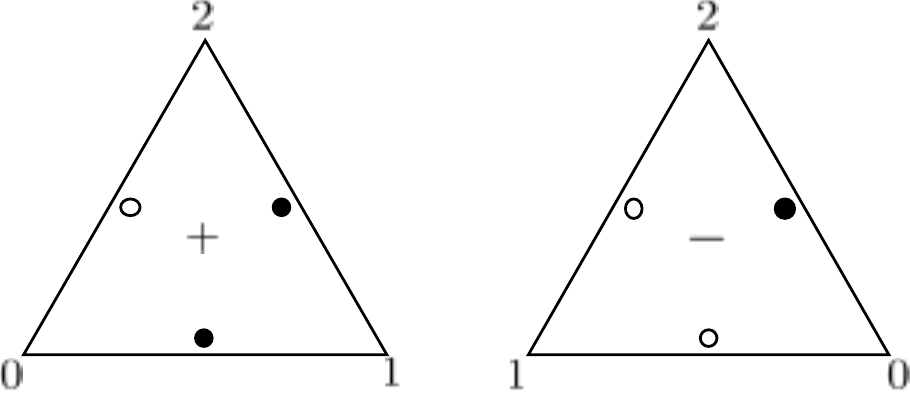}
\caption{Assignment of Grassmann variables on 1-simplices. $\theta$ (resp.~$\overline{\theta}$) is represented as a black (resp.~white) dot.}
\label{fig:Grassmann}
\end{figure}

Due to the fermionic sign of Grassmann variables, $\sigma(\alpha)$ becomes a quadratic function, whose quadratic property depends on the order of Grassmann variables in $u(t)$. We will adopt the order used in Gaiotto-Kapustin \cite{Gaiotto:2015zta}, which is defined as  $u(012)=\vartheta_{12}^{\alpha(12)}\vartheta_{01}^{\alpha(01)}\vartheta_{02}^{\alpha(02)}$ when $(012)$ is a $+$ triangle, 
and $u(012)=\vartheta_{02}^{\alpha(02)}\vartheta_{01}^{\alpha(01)}\vartheta_{12}^{\alpha(12)}$ for a $-$ triangle. 
We choose the assignment of $\theta$ and $\overline{\theta}$ on each $e$ in the following fashion: 
the Grassmann variables on $e$ are assigned such that, if $t$ is a $+$ (resp.~$-$) simplex, $u(t)$ includes $\overline{\theta}_e$ when $e$ is given by omitting a vertex with odd (resp.~even) number from $t=(012)$, see Fig.~\ref{fig:Grassmann}.

Based on the above definition of $u(t)$, the quadratic property of $\sigma(\alpha)$ turns out to be
\begin{equation}
    \sigma(\alpha)\sigma(\alpha')=\sigma(\alpha+\alpha')(-1)^{\int\alpha\cup\alpha'},
    \label{eq:ClosedQuad}
\end{equation}
for closed $\alpha, \alpha'$.
Moreover, the change of $\sigma(\alpha)$ under the gauge transformation
$\alpha\to \alpha+\delta \gamma$ or under the change of the triangulation is controlled by the formula
\begin{equation}
\sigma(\widetilde M,\widetilde \alpha) 
= (-1)^{\int_K w_2 \cup \alpha} \sigma(M,\alpha), 
\label{eq:WZWspin}
\end{equation}
where $\widetilde M$ is the same manifold $M$ with a different triangulation,
$\widetilde\alpha$ is a cocycle such that $[\alpha]=[\widetilde\alpha] $ in cohomology,
and $K=M\times [0,1]$ such that the two boundaries are given by $M$ and $\widetilde M$,
and finally $\alpha$ is extended to $K$ so that it restricts to $\alpha$ and $\widetilde \alpha$ on the boundaries. The derivation of~\eqref{eq:ClosedQuad},~\eqref{eq:WZWspin} was given in \cite{Gaiotto:2015zta}.
Then, the spin theory $z[M,\eta, \alpha]$ is defined as
\begin{equation}
z[M, \eta,\alpha]=\sigma(M, \alpha)(-1)^{\int_M\eta\cup\alpha},
\label{eq:spintft2}
\end{equation}
where $\eta$ specifies a spin structure on $M$, 
and satisfies $\delta\eta=w_2$.

\subsection{pin TQFT on the lattice}
\label{subsec:pintft}
We construct an unoriented manifold by picking locally oriented patches, and then gluing them along codimension one loci by transition functions. The locus where the transition functions are orientation reversing, constitutes a representative of the dual of the first Stiefel-Whitney class $w_1$. We will sometimes call the locus an orientation reversing wall.
We can choose a consistent orientation everywhere if we remove a locus of the orientation reversing wall.

\begin{figure}[htb]
\centering
\includegraphics{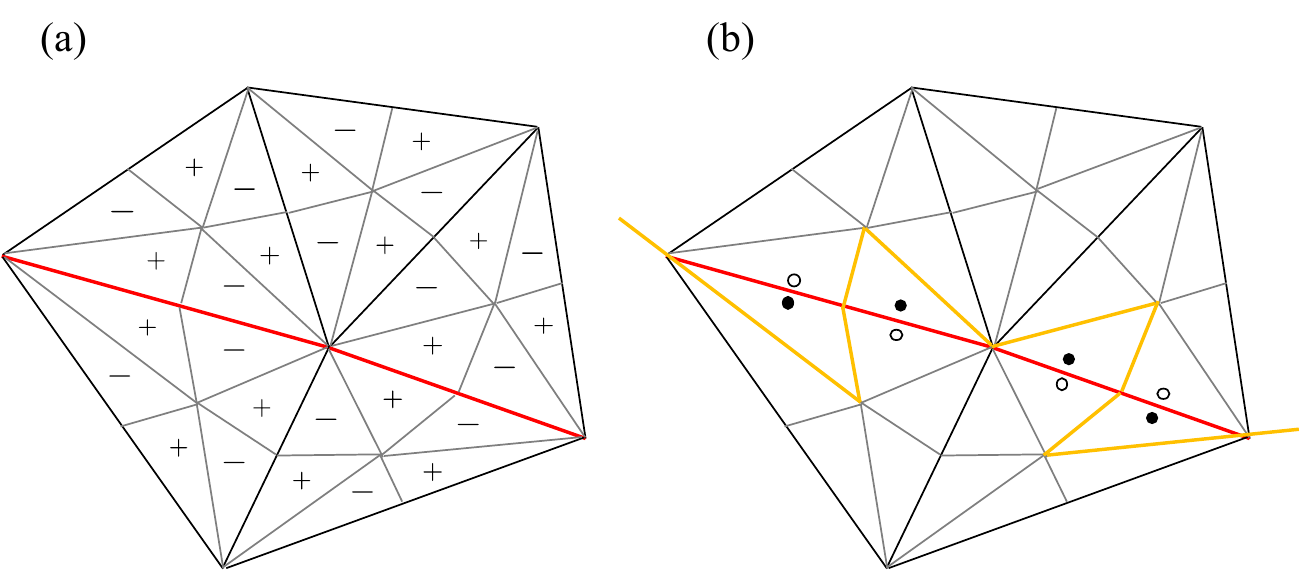}
\caption{(a): The signs of triangles near the orientation reversing wall, which is represented as a red line. (b): Assignment of Grassmann variables on the wall specifies a deformation of the wall that intersects the wall transversally at vertices.
}
\label{fig:wall}
\end{figure}

We remark that the assigning rule of the Grassmann variables described in the previous subsection fails, when $e$ lies on the wall where we glue patches of $M$ by the orientation reversing map. In this case, we would have to assign Grassmann variables of the same color on both sides of $e$ (i.e., both are black ($\theta$) or white ($\overline{\theta}$)), since the two triangles sharing $e$ have the identical sign when $e$ is on the orientation reversing wall, see Fig.~\ref{fig:wall} (a). Hence, we need to slightly modify the construction of the Grassmann integral on the orientation reversing wall. To do this, instead of specifying a canonical rule to assign Grassmann variables on the wall,
we just place a pair $\theta_e$, $\overline{\theta}_{e}$ on the wall in an arbitrary fashion. 

Along with this modification, the Grassmann integral on $M$ is revised as
\begin{equation}
    \sigma(M, \alpha)=\int\prod_{e|\alpha(e)=1}d\theta_e d\overline{\theta}_e \prod_t u(t)\prod_{e|\mathrm{wall}}(\pm i)^{\alpha(e)},
    \label{sigmadefpin}
\end{equation}
where the $\prod_{e|\mathrm{wall}}(\pm i)^{\alpha(e)}$ term assigns weight $(+i)^{\alpha(e)}$ (resp.~$(-i)^{\alpha(e)}$) on each 1-simplex $e$ on the orientation reversing wall, when $e$ is shared with $+$ (resp.~$-$) 2-simplices. There is no ambiguity in such definition, since both 2-simplices on the side of $e$ have the same sign when $e$ is on the wall.

The quadratic property of the Grassmann integral \eqref{eq:ClosedQuad} still holds for the pin$^{\pm}$ case while the effect of re-triangulations and gauge transformations are given by 
\begin{equation}
\sigma(\widetilde M,\widetilde \alpha) 
= (-1)^{\int_K (w_2+w_1^2) \cup \alpha} \sigma(M,\alpha), 
\label{eq:WZW}
\end{equation}
as shown in \cite{Kobayashi2019}. 

Then, the pin$^{\pm}$ theory $z[M,\eta, \alpha]$ is defined as
\begin{equation}
z[M, \eta,\alpha]=\sigma(M, \alpha)(-1)^{\int_M\eta\cup\alpha},
\label{eq:pintft2}
\end{equation}
where $\eta$ specifies a pin$^{\pm}$ structure on $M$, which satisfies $\delta\eta=w_2$ (resp.~$\delta\eta=w_2+w_1^2$) in the pin$^+$ (resp.~pin$^-$) case.

Here, it should be emphasized that the expressions~\eqref{eq:WZW},~\eqref{eq:pintft2} are based on a specific choice of the representative of the Poincar\'{e} dual of $w_2$, $w_1^2$ in $M$. Firstly, the representative of the dual of $w_2$ on $M$ is given by the set of all vertices of the barycentric subdivision~\cite{MS, HalperinToledo, BlantonMcCrory}. Secondly, to specify the dual of $w_1^2$ on $M$, we first observe that the choice of the assignment of Grassmann variables on the wall corresponds to choosing the slight deformation of the wall, such that the deformation intersects transversally with the wall at vertices. Concretely, we deform the wall on each edge of the wall to the side where $\theta$ (black dot) is contained, see Fig.~\ref{fig:wall} (b). Here, both walls before and after deformation give a representative of the dual of $w_1$, and then the intersection of two walls gives our representative of the dual of $w_1^2$. $\eta$ in~\eqref{eq:pintft2} is a trivialization of the representative of the obstruction class, prepared in the above fashion.

\if
We endow $M$ with a triangulation. In addition, we take the barycentric subdivision for the triangulation of $M$. Namely, each 2-simplex in the initial triangulation of $M$ is subdivided into 6 simplices, whose vertices are barycenters of the subsets of vertices in the 2-simplex. We further assign a local ordering to vertices of the barycentric subdivision, such that a vertex on the barycenter of $i$ vertices is labeled as $i$.
\fi
\if
Each simplex can then be either a $+$ simplex or a $-$ simplex, depending on whether the ordering agrees with the local orientation or not.
We assign a pair of Grassmann variables $\theta_e, \overline{\theta}_e$ on each 1-simplex $e$ of $M$ such that $\alpha(e)=1$, we associate $\theta_e$ on one side of $e$ contained in one of 2-simplices neighboring $e$ (which will be specified later), $\overline{\theta}_e$ on the other side. Then, $\sigma(M, \alpha)$ is defined as
\begin{equation}
    \sigma(M, \alpha)=\int\prod_{e|\alpha(e)=1}d\theta_e d\overline{\theta}_e \prod_t u(t)\prod_{e|\mathrm{wall}}(\pm i)^{\alpha(e)},
    \label{sigmadefpin}
\end{equation}
where $t$ denotes a 2-simplex, and $u(t)$ is the product of Grassmann variables contained in $t$.
Namely, $u(t)$ on $t=(012)$ is the product of
$\vartheta_{12}^{\alpha(12)}, \vartheta_{01}^{\alpha(01)}, \vartheta_{02}^{\alpha(02)}$. 
Here, $\vartheta$ denotes $\theta$ or $\overline{\theta}$ depending on the choice of the assigning rule, which will be discussed later. The order of Grassmann variables in $u(t)$ will also be defined shortly.
We note that $u(t)$ is ensured to be Grassmann-even when $\alpha$ is closed. 
The $\prod_{e|\mathrm{wall}}(\pm i)^{\alpha(e)}$ term assigns weight $(+i)^{\alpha(e)}$ (resp.~$(-i)^{\alpha(e)}$) on each 1-simplex $e$ on the orientation reversing wall, when $e$ is shared with $+$ (resp.~$-$) 2-simplices. There is no ambiguity in such definition, since both 2-simplices on the side of $e$ have the same sign when $e$ is on the wall.
\fi
\if
Due to the fermionic sign of Grassmann variables, $\sigma(\alpha)$ becomes a quadratic function, whose quadratic property depends on the order of Grassmann variables in $u(t)$. We will adopt the order used in Gaiotto-Kapustin \cite{Gaiotto:2015zta}, which is defined as  $u(012)=\vartheta_{12}^{\alpha(12)}\vartheta_{01}^{\alpha(01)}\vartheta_{02}^{\alpha(02)}$ when $(012)$ is a $+$ triangle, 
and $u(012)=\vartheta_{02}^{\alpha(02)}\vartheta_{01}^{\alpha(01)}\vartheta_{12}^{\alpha(12)}$ for a $-$ triangle. 
We choose the assignment of $\theta$ and $\overline{\theta}$ on each $e$ in the following fashion. First, let us consider the case when $e$ is not contained in the orientation reversing wall. Then, the Grassmann variables on $e$ are assigned such that, if $t$ is a $+$ (resp.~$-$) simplex, $u(t)$ includes $\overline{\theta}_e$ when $e$ is given by omitting a vertex with odd (resp.~even) number from $t=(012)$, see Fig.~\ref{fig:Grassmann}.
\fi


\if
However, we remark that the assigning rule fails, when $e$ lies on the wall where we glue patches of $M$ by the orientation reversing map. In this case, we would have to assign Grassmann variables of the same color on both sides of $e$ (i.e., both are black ($\theta$) or white ($\overline{\theta}$)), since the two triangles sharing $e$ have the identical sign when $e$ is on the orientation reversing wall, see Fig.~\ref{fig:wall} (a). Hence, we need to slightly modify the construction of the Grassmann integral on the orientation reversing wall. To do this, instead of specifying a canonical rule to assign Grassmann variables on the wall,
we just place a pair $\theta_e$, $\overline{\theta}_{e}$ on the wall in an arbitrary fashion. 
\fi


\if
Based on the above definition of $u(t)$, the quadratic property of $u(t)$ turns out to be
\begin{equation}
    \sigma(\alpha)\sigma(\alpha')=\sigma(\alpha+\alpha')(-1)^{\int\alpha\cup\alpha'},
    \label{eq:ClosedQuad}
\end{equation}
for closed $\alpha, \alpha'$.
Moreover, the change of $\sigma(\alpha)$ under the gauge transformation
$\alpha\to \alpha+\delta \gamma$ or under the change of the triangulation is controlled by the formula
\begin{equation}
\sigma(\widetilde M,\widetilde \alpha) 
= (-1)^{\int_K (w_2+w_1^2) \cup \alpha} \sigma(M,\alpha), 
\label{eq:WZW}
\end{equation}
where $\widetilde M$ is the same manifold $M$ with a different triangulation,
$\widetilde\alpha$ is a cocycle such that $[\alpha]=[\widetilde\alpha] $ in cohomology,
and $K=M\times [0,1]$ such that the two boundaries are given by $M$ and $\widetilde M$,
and finally $\alpha$ is extended to $K$ so that it restricts to $\alpha$ and $\widetilde \alpha$ on the boundaries. The derivation of~\eqref{eq:ClosedQuad},~\eqref{eq:WZW} was given in \cite{Kobayashi2019}.
Then, the pin$^{\pm}$ theory $z[M,\eta, \alpha]$ is defined as
\begin{equation}
z[M, \eta,\alpha]=\sigma(M, \alpha)(-1)^{\int_M\eta\cup\alpha},
\label{eq:pintft2}
\end{equation}
where $\eta$ specifies a pin$^{\pm}$ structure of $M$, which satisfies $\delta\eta=w_2$ (resp.~$\delta\eta=w_2+w_1^2$) in the pin$^+$ (resp.~pin$^-$) case.
\fi
\if
Here, it should be emphasized that the expressions~\eqref{eq:WZW},~\eqref{eq:pintft2} are based on a specific choice of the representative of the Poincar\'{e} dual of $w_2$, $w_1^2$ in $M$. Firstly, the representative of the dual of $w_2$ on $M$ is given by the set of all vertices of the barycentric subdivision. Secondly, to specify the dual of $w_1^2$ on $M$, first we observe that the choice of the assignment of Grassmann variables on the wall corresponds to choosing the slight deformation of the wall, such that the deformation intersects transversally with the wall at vertices. Concretely, we deform the wall on each edge of the wall to the side where $\theta$ (black dot) is contained, see Fig.~\ref{fig:wall} (b). Here, both walls before and after deformation give a representative of the dual of $w_1$, and then the intersection of two walls gives our representative of the dual of $w_1^2$. $\eta$ in~\eqref{eq:pintft2} is a trivialization of the representative of the obstruction class, prepared in the above fashion.
\fi

\subsection{Arf-Brown-Kervaire invariant in (1+1)d}
\label{subsec:ABK}
In this subsection, we construct the 2d pin$^{-}$ invertible TQFT~\cite{Debray2018} for the
Arf-Brown-Kervaire (ABK) invariant via the Grassmann integral on the lattice,  whose state sum definition was initially given in~\cite{Turzillo2018}. 
In condensed matter literature, this invertible theory describes
(1+1)d topological superconductors
in class BDI.
Here, we construct the $\bZ_8$-valued ABK invariant by coupling the 2d state sum shadow TQFT with the Grassmann integral.
For  the $\bZ_2$-valued Arf invariant of the spin case,
this was done in ~\cite{Gaiotto:2015zta}.

The weight for the state sum is assigned in the same manner as the case of the Arf invariant of the spin case~\cite{Gaiotto:2015zta}, described as follows. 
For a given configuration
$\alpha\in C^1(M, \bZ_2)$, we assign weight $1/2$ to each 1-simplex $e$, and also assign weight $2$ to each 2-simplex $f$ when $\delta\alpha=0$ at $f$, otherwise 0. Let us denote the product of the whole weight as $\widetilde{Z}[\alpha]$.
Then, we can see that the partition function is given by the ABK invariant up to the Euler term, 
\begin{equation}
\begin{split}
    Z[M,\eta]&=\sum_{\alpha\in Z^1(M,\bZ_2)}\sigma(M,\alpha)(-1)^{\int_M \eta\cup\alpha}\widetilde{Z}[\alpha] \\
    &=2^{|F|-|E|}\cdot\sum_{\alpha\in Z^1(M,\bZ_2)}\sigma(M,\alpha)(-1)^{\int_M \eta\cup\alpha}\\
    &=2^{\chi(M)-1}\cdot\sum_{[\alpha]\in H^1(M,\bZ_2)}\sigma(M,\alpha)(-1)^{\int_M \eta\cup\alpha}\\
    &=\sqrt{2}^{\chi(M)}\mathrm{ABK}[M,\eta],
    \end{split}
    \label{eq:abk}
\end{equation}
where $|F|$, $|E|$ denotes the number of 2-simplices, 1-simplices in $M$, respectively. $\chi(M)$ denotes the Euler characteristic of $M$, and ABK$[M,\eta]$ is the
ABK invariant,
\begin{align}
    \mathrm{ABK}[M,\eta]=\frac{1}{\sqrt{|H^1(M,\bZ_2)|}}\sum_{[\alpha]\in H^1(M,\bZ_2)}i^{Q_{\eta}[\alpha]}.
\end{align}
Here, $i^{Q_{\eta}[\alpha]}=\sigma(M,\alpha)(-1)^{\int_M \eta\cup\alpha}$ is a $\bZ_4$-valued quadratic function that satisfies~\cite{BFquadratic}
\begin{equation}
    Q_{\eta}[\alpha]+Q_{\eta}[\alpha']=Q_{\eta}[\alpha+\alpha']+2\int_M\alpha\cup\alpha'.
    \label{eq:refine}
\end{equation}

The ABK invariant determines the pin$^-$ bordism class of 2d manifolds $\Omega_2^{\mathrm{pin}^-}(\mathrm{pt})=\bZ_8$, which is generated by $\mathbb{RP}^2$~\cite{KirbyTaylor}. 
To see this, let $\alpha$ be
a nontrivial 1-cocycle that generates $H^1(\mathbb{RP}^2, \bZ_2)=\bZ_2$. Then, using the quadratic property for $\alpha=\alpha'$ in~\eqref{eq:refine}, one can see that $Q_{\eta}[\alpha]$ takes value in $\pm 1$, since $Q_{\eta}[0]=0$ and $\int_M \alpha\cup\alpha'=1$.  
$Q_{\eta}[\alpha]=\pm 1$ corresponds to two possible pin$^{-}$ structures on $\mathbb{RP}^2$. 
Then, the ABK invariant is given by an 8th root of unity,
\begin{equation}
\mathrm{ABK}[M, \eta]=\frac{1\pm i}{\sqrt{2}}=e^{\pm 2\pi i/8}.
\end{equation}
If $M$ is oriented, the ABK invariant reduces to the Arf invariant
$\mathrm{Arf}[M,\eta]$,
which determines the spin bordism class $\Omega_2^{\mathrm{spin}}(\mathrm{pt})=\mathbb{Z}_2$~\cite{SpinBordismManifoldAtlas}.

\subsection{Wave function on boundaries}
\label{subsec:wfn}
Now let us consider a spin or pin TQFT on $M$ constructed in the manner described in Sec.~\ref{subsec:spintft} and \ref{subsec:pintft}, when $M$ has a non-empty boundary. 

To construct the wave function of the vacuum state, let us describe the state of spin/pin TQFT on the lattice.
We recall that the state-sum model of 2d oriented spin TQFT is built from a $\mathbb{Z}_2$-graded (not necessarily commutative) semi-simple Frobenius algebra $A$~\cite{Gaiotto:2015zta, Fukuma-Hosono-Kawai1994}. 
 A similar construction for the spin TQFT wave functions
is also found in~\cite{Kapustin-Turzillo-You2018}.
If one seeks to consider unoriented pin case, we further have to assume that $A$ is commutative, to ensure invariance of the theory under re-triangulation~\cite{Karimipour1995}.

Let $\epsilon_i\in A (i\in I)$ be the basis of $A$
and denote $\alpha(i)$ as the $\mathbb{Z}_2$ grading of $\epsilon_i$. We sometimes call the $\mathbb{Z}_2$ grading as the fermion parity.
We write $C^i_{jk}$ as the structure constant of $A$ in this basis. Then, let $g_{ij}:=C^l_{ik}C^k_{jl}$. Because $g_{ij}$ is non-degenerate, it has the inverse denoted as $g^{ij}$. We define $C_{ijk}:=g_{il}C^l_{jk}$, which turns out to be cyclically symmetric. If we further assume that $A$ is commutative, $C_{ijk}$ is symmetric under any permutations.
Since the algebra $A$ is $\mathbb{Z}_2$-graded, $C_{ijk}$ and $g^{ij}$ always respect the $\mathbb{Z}_2$ grading. Namely, $C_{ijk}$ vanishes unless $\alpha(i)+\alpha(j)+\alpha(k)\equiv 0$
(mod 2), and $g^{ij}$ vanishes unless $\alpha(i)+\alpha(j)\equiv 0$
(mod 2).

Using these data, we can construct the bosonic shadow theory $\widetilde{Z}_M[\alpha]$ coupled with background $\mathbb{Z}_2$ gauge field $\alpha$.
We assign a pair of elements $\epsilon_i, \epsilon_j$ of $A$ on each 1-simplex of $M$. 
Here, the background field $\alpha\in Z^1(M, \mathbb{Z}_2)$ is regarded as the $\mathbb{Z}_2$ grading of the element in $A$ assigned on 1-simplices. Namely, a pair of elements share the same $\mathbb{Z}_2$ grading specified by $\alpha$.
Then, we assign weight $g^{ij}$ on each 1-simplex and $C_{ijk}$ on each 2-simplex.
To obtain the partition function $\widetilde{Z}_M[\alpha]$, we just have to perform contraction of indices for all factors $g^{ij},C_{ijk}$ on $M$ with a fixed $\mathbb{Z}_2$ grading $\alpha$. Then, the spin/pin TQFT is constructed in the form of~\eqref{eq:gaugedpin}, by coupling with the spin/pin theory prepared by the Grassmann integral.

If one makes up a boundary on $M$, the Hilbert space for the TQFT is constructed on $A^{\widehat{\otimes} n}$, where $n$ is the number of boundary 1-simplices.
Here, $\widehat{\otimes}$ denotes the supertensor product of $\mathbb{Z}_2$-graded algebra. 
Compared with conventional tensor product, supertensor product is modified in the way which respects the fermionic sign of the elements carrying fermion parity. Namely, given the algebra $A_1\ \widehat{\otimes}\ A_2$, the multiplication of elements in $A_1\ \widehat{\otimes}\ A_2$ is defined as
\cite{Kapustin-Turzillo-You2018}
\begin{align}
    (\epsilon_1 \ \widehat{\otimes}\ \epsilon_2) \cdot (\epsilon'_1 \ \widehat{\otimes}\ \epsilon'_2) = (-1)^{\alpha(\epsilon_2)\alpha(\epsilon'_1)}\cdot (\epsilon_1\cdot\epsilon'_1)\  \widehat{\otimes}\ (\epsilon_2\cdot\epsilon'_2).
\end{align}

For a given basis of $A^{\widehat{\otimes} n}$ on $\partial M$, the wave function is evaluated as the path integral on $M$.
Denoting the element of $A^{\widehat{\otimes}n}$ in the form of $\ket{\epsilon_{1}\dots \epsilon_{n}}$, the wave function for the prepared Hilbert space is given by evaluating the path integral of the TQFT~\eqref{eq:pintft} on $M$, which is expressed as
\begin{align}
    \ket{\psi} = \sum_{\alpha \in Z^1({M}, \mathbb{Z}_2)}\sum_{A^{\widehat{\otimes}n}} \widetilde{Z}_{M}[\alpha] \sigma(M, \alpha; \mathrm{ord}(1,\cdots,n)) (-1)^{\int_M \eta \cup \alpha} \ket{\epsilon_{1}\dots \epsilon_{n}}.
\end{align}

Especially, let us consider the simplest case where $A=Cl(1)$ (real Clifford algebra generated by one $\mathbb{Z}_2$-odd element), 
where we are setting $g^{jk}=1/2\cdot \delta_{jk}, C_{ijk}=2\delta_{i+j,k}$. 
For simplicity, let $M$ be an oriented spin surface. Then, we can build the Hilbert space on $\partial M$ as the Fock space of $n$ complex fermions.
Namely, we prepare a complex fermion on each boundary 1-simplex,
and consider a Fock space of the fermions. Then, the wave function for the prepared Hilbert space is given by evaluating the path integral of the TQFT~\eqref{eq:pintft} on $M$, which is expressed as
\begin{align}
\ket{\psi} = \sum_{\alpha \in Z^1({M}, \mathbb{Z}_2)} \widetilde{Z}_{M}[\alpha] \sigma(M, \alpha; \mathrm{ord}(1,\cdots, n)) (-1)^{\int_M \eta \cup \alpha} (c^{\dagger}_{1})^{\alpha(e_{1})} \cdots (c^{\dagger}_{n})^{\alpha(e_n)} \ket{0},
\label{eq:gs}
\end{align}
where $c_j/c^{\dagger}_j$ denotes a complex fermion 
annihilation/creation operator at $e_j$. 
$\widetilde{Z}_M[\alpha]$ is the weight of the bosonic shadow theory evaluated on $M$.
$\sigma(M, \alpha; \mathrm{ord}(1,\cdots,n))$ evaluates the Grassmann integral on an open surface $M$, which is defined via the following relation
\begin{equation}
\int_M \prod_{e|\alpha(e) = 1} d\theta_e d\overline{\theta}_e \prod_{t} u(t) 
= \sigma(M, \alpha; \mathrm{ord}(1, \cdots, n)) \vartheta_{1}^{\alpha(e_1)} \cdots \vartheta_{n}^{\alpha(e_n)}.
\label{eq:openGrassmann}
\end{equation}
Here, $\vartheta_{j}$ represents $\theta_{j}$ or $\overline{\theta}_{j}$ depending on an assignment of Grassmann variables on boundaries.

In the expression~\eqref{eq:gs}, $\eta$ satisfies $\delta\eta=w_2$ as an element of $Z^2(M,\partial M; \mathbb{Z}_2)$, where the representative of $w_2$ is specified as the dual of a set of all 0-simplices in the barycentric subdivision, as illustrated in Sec.~\ref{subsec:pintft}. 
Thus, we can rewrite the factor $(-1)^{\int_M \eta\cup\alpha}$ as
\begin{equation}
(-1)^{\int_M \eta \cup \alpha} = (-1)^{\int_{E_M} \alpha},
\label{eq:etadual}
\end{equation}
where $E_M$ is the dual of $\eta$, 
and
$\partial E_M$ becomes a set of all 0-simplices in the barycentric subdivision, when restricted to the interior of $M$.

The algebra $A = Cl(1)$ also works as data for the unoriented pin$^{-}$ TQFT presented in Sec.~\ref{subsec:ABK}. When $M$ is a pin$^-$ surface, we include the $\mathbb{Z}_4$ factor $\prod_{e|\mathrm{wall}} (\pm i)^{\alpha(e)}$ in the lhs of the equation \eqref{eq:openGrassmann}, and let $\eta$ be a trivialization of $w_2+w_1^2$ as an element of $Z^2(M,\partial M; \mathbb{Z}_2)$.

\subsection{Ground state wave function of the Kitaev chain}
\label{subsec:kitaev}
Here, we provide the 
fixed-point ground state wave function of topological superconductors in class BDI, based on the 2d pin$^-$ TQFT described in Sec.~\ref{subsec:ABK}.
In this case, the shadow theory on a closed spin or pin$^{-}$ manifold is given by
\begin{equation}
\widetilde{Z}_X[\alpha] = 2^{|F|-|E|},
\end{equation}
where $|F|$ and $|E|$ denote the number of faces and edges of $X$ respectively.
We set the shadow theory $\widetilde{Z}_M$ on $M$ with a non-empty boundary, by requiring that the wave function $\ket{\mathrm{in}}$ is correctly normalized,
\begin{equation}
\braket{\mathrm{out} | \mathrm{in}} = 1.
\end{equation}
where $\bra{\mathrm{out}}$ is conjugate to $\ket{\mathrm{in}}$. 
From this condition, the shadow theory is obtained as
\begin{equation}
\widetilde{Z}_M[\alpha] = 2^{|F| - |E| +|E_b| / 2 - \chi(Y) / 4},
\label{eq: shadow M}
\end{equation}
where $|E_b|$ denotes the number of boundary 1-simplices and $\chi(Y)$ is the Euler characteristic of $Y$ which is obtained by gluing $M$ and $\overline{M}$ along the boundary. 
Actually, the product of the shadow theories on $M$ and $\overline{M}$ 
\begin{equation}
\widetilde{Z}_M[\alpha|_M] \widetilde{Z}_{\overline{M}}[\alpha|_{\overline{M}}] = 2^{-\chi(Y) / 2} \widetilde{Z}_{Y}[\alpha]
\end{equation}
gives the shadow theory on $Y$,
and the norm of the wave function is 
\begin{equation}
\begin{aligned}
\braket{\mathrm{out} | \mathrm{in}} & = 2^{-\chi(Y) / 2} \sum_{\alpha \in Z^1(Y, \mathbb{Z}_2)} \widetilde{Z}_{Y}[\alpha] \sigma(Y, \alpha) (-1)^{\int_{Y} \eta \cup \alpha}\\
& = \mathrm{ABK}[Y, \eta].
\end{aligned}
\end{equation}
Here, ABK$[Y, \eta]$ is the Arf-Brown-Kervaire invariant quantized as the 8th root of unity. 
Since the above expression is positive (ensured by reflection positivity of unitary TQFT~\cite{Freed:2016rqq, Yonekura:2018ufj}), ABK$[Y, \eta]$ is 1, which shows that the wave function is correctly normalized.

\if
In the same way, one can compute the many-body topological invariants 
\begin{equation}
\begin{aligned}
& \braket{\mathrm{out} | \mathcal{R}_{\mathrm{part}} | \mathrm{in}} = 2^{-\chi(X_0) / 2 + \chi(X) / 2} \mathrm{ABK}[X, \eta]\\
& \mathrm{tr} (\rho_I U_{I_1} \rho_I^{T_1} U_{I_1}^{\dagger}) = 2^{-\chi(X_0) + \chi(X) / 2} \mathrm{ABK}[X, \eta]
\end{aligned}
\end{equation}
In particular, if we choose $M$ as a disk, $X_0 = S^2$ and $X = \mathbb{R}P^2$, which gives 
\begin{equation}
\begin{aligned}
& \braket{\mathrm{out} | \mathcal{R}_{\mathrm{part}} | \mathrm{in}} = \frac{1}{\sqrt{2}} e^{\pm 2\pi i /8}\\
& \mathrm{tr} (\rho_I U_{I_1} \rho_I^{T_1} U_{I_1}^{\dagger}) = \frac{1}{2\sqrt{2}} e^{\pm 2\pi i / 8}
\end{aligned}
\end{equation}
This reproduces the results obtained in \cite{Shapourian-Shiozaki-Ryu2017a}. 
\fi

\section{Partial time-reversal}
\label{sec:pt}
In this section, we will formulate the quantized non-local order parameter for SPT phases proposed in~\cite{Shapourian-Shiozaki-Ryu2017a}, for the states prepared by spin or pin TQFT.
First, we recall the construction of the order parameter for (1+1)d SPT phases in class BDI, following~\cite{Shapourian-Shiozaki-Ryu2017a}.

Let us consider a ground state of the SPT phase on a ring with length $n$, constructed in the Fock space of complex fermions $c_1,\dots, c_n$
with anti-periodic boundary condition.
We take the reduced density matrix $\rho_I$ of the ground state, defined on an interval $I$ in the ring. Then, we take a bipartition of $I$ as $I=I_1\sqcup I_2$. Roughly speaking, the order parameter is defined via the process of taking the ``transpose'' of the density matrix, restricted to the interval $I_1$. With a proper definition of the transpose in the partial region $I_1$ in $I$, the order parameter is given by
\begin{align}
    \mathrm{tr}_{I}(\rho_I \rho_{I}^{\mathcal{T}_1}),
    \label{eq:ptr}
\end{align}
where $\rho_{I}^{\mathcal{T}_1}$ denotes the density matrix acted by ``partial time-reversal''.
The definition of partial time-reversal
is transparently expressed in the coherent state basis. Namely, we introduce $n$ Grassmann variables $\xi_1, \dots, \xi_n$ and denote a state like $\ket{\{\xi_i\}}=\prod_j\exp(-\xi_j c^{\dagger}_j)\ket{0}$. 
The density matrix is rewritten in the coherent state basis as
\begin{align}
    \rho_I=\int d[\overline{\xi}, \xi]d[\overline{\chi}, \chi]\ket{\{\xi_j\}}\rho_I(\{\overline{\xi}_j\}; \{\chi_j\})\bra{\{\overline{\chi}_j\}},
\end{align}
where $d[\overline{\xi}, \xi]=\prod_j d\overline{\xi}_j d\xi_j e^{-\sum_j\overline{\xi}_j\xi_j}$, and $\rho_I(\{\overline{\xi}_j\}; \{\chi_j\})=\bra{\{\overline{\xi}_j\}}\rho_I\ket{\{\chi_j\}}$. Then, the operation $\rho_{I}^{\mathcal{T}_1}$ is defined as
\begin{align}
    \rho_{I}^{\mathcal{T}_1}:=\int d[\overline{\xi}, \xi]d[\overline{\chi}, \chi]\ket{\{i\overline{\chi}_j\}_{j\in I_1}, \{\xi_j\}_{j\in I_2}}\rho_I(\{\overline{\xi}_j\}; \{\chi_j\})\bra{\{i\xi_j\}_{j\in I_1}, \{\overline{\chi}_j\}_{j\in I_2}}.
    \label{eq:ptrans}
\end{align}
This operation on $I_1$ is called partial time-reversal in~\cite{Shapourian-Shiozaki-Ryu2017a}, 
since 
it acts on Grassmann variables in $I_1$
in the same fashion as the time-reversal for the symmetry class BDI.

In the following, we formulate partial time-reversal and compute the quantity~\eqref{eq:ptr}, on a wave function constructed from a (1+1)d pin$^{-}$ invertible TQFT discussed in Sec.~\ref{subsec:wfn}. 
We find that~\eqref{eq:ptr} is identical to the partition function of the pin$^-$ TQFT $Z[X,\eta]$ evaluated on a closed unoriented pin$^{-}$ surface $X$, which generates the pin$^-$ bordism group $\Omega_2^{\mathrm{pin}^{-}}(\mathrm{pt})=\mathbb{Z}_8$. Especially, when $M$ is taken to be a disk, $X=\mathbb{RP}^2$ and
\begin{align}
    \mathrm{tr}_{I}(\rho_I \rho_{I}^{\mathcal{T}_1})=Z[\mathbb{RP}^2, \eta],
    \label{eq:ptRP2}
\end{align}
where $\eta$ specifies a pin$^-$ structure.
The partial time-reversal can also be defined on the Hilbert space of spin/pin TQFT prepared by $\mathbb{Z}_2$-graded Frobenius algebra, which is described in~\ref{app:ptrfrob}. Though we will mostly work on the Kitaev chain wave function in the main text, the correspondence between the quantized non-local order parameter and TQFT wave function~\eqref{eq:ptRP2} is safely extended to a pin$^-$ TQFT prepared by generic commutative $\mathbb{Z}_2$-graded Frobenius algebra.

\subsection{Evaluation of partial time-reversal}
\label{subsec:ptrcomp}
\begin{figure}[htb]
\centering
\includegraphics{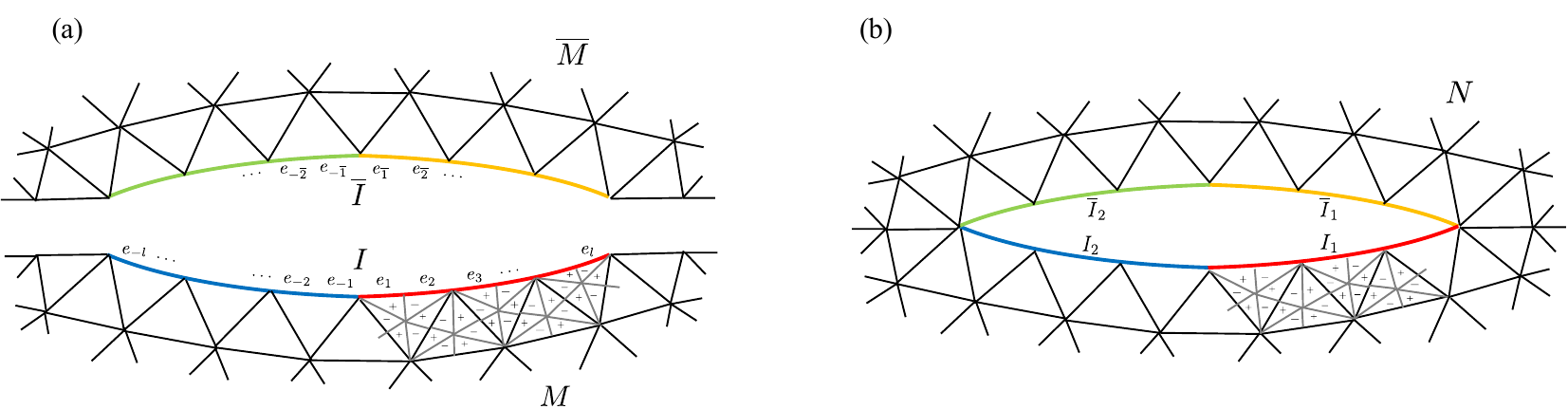}
\caption{(a): $\ket{\mathrm{in}}$ and $\bra{\mathrm{out}}$ are prepared by the path integral on $M$ and $\overline{M}$ respectively. (b): Taking the partial trace amounts to gluing $M$ and $\overline{M}$, and the resulting surface is denoted as $N$.
}
\label{fig:MMN}
\end{figure}

Now we perform the explicit computation of~\eqref{eq:ptr}. We start with constructing the reduced density matrix. We first prepare the state on $\partial M=S^1$ and its conjugation, in the form of~\eqref{eq:gs}
\begin{align}
& \ket{\mathrm{in}} = \sum_{\alpha \in Z^1({M}, \mathbb{Z}_2)} \widetilde{Z}_{M}[\alpha] \sigma(M, \alpha; \mathrm{ord}(-n,\cdots,n)) (-1)^{\int_M \eta \cup \alpha} (c^{\dagger}_{-n})^{\alpha(e_{-n})} \cdots (c^{\dagger}_{n})^{\alpha(e_n)} \ket{0},
\label{eq:in}\\
& \bra{\mathrm{out}} = \sum_{\alpha \in Z^1({\overline{M}}, \mathbb{Z}_2)} \widetilde{Z}_{\overline{M}}[\alpha] \sigma(\overline{M}, \alpha; \mathrm{ord}(\overline{n}, \cdots, -\overline{n})) (-1)^{\int_{\overline{M}} \eta \cup \alpha} \bra{0} c_{\overline{n}}^{\alpha(e_{\overline{n}})} \cdots c_{-\overline{n}}^{\alpha(e_{-\overline{n}})},
\label{eq:out}
\end{align}
where we let the number of boundary 1-simplices $2n$ here, and labeled 1-simplices in $\partial M$ as $e_{-n},\dots, e_{-1}, e_{1}, \dots, e_n$, for later convenience.
$\overline{M}$ is given by reversing the orientation of $M$, and we denote 1-simplices in $\partial \overline{M}$ as $e_{-\overline{n}}, \dots, e_{-\overline{1}}, e_{\overline{1}},\dots e_{\overline{n}}$.
Starting from the density matrix $\rho=\ket{\mathrm{in}} \bra{\mathrm{out}}$, we take the reduced density matrix $\rho_I$ for the interval $I=\sum_{1 \leq |j| \leq l} e_j$, see Fig.~\ref{fig:MMN}. For simplicity, we set $l, n$ as even. Then, $\rho_I$ is expressed as
\begin{equation}
\begin{aligned}
\rho_I = & \sum_{\alpha \in Z^1(N, \mathbb{Z}_2)} \widetilde{Z}_N[\alpha] \sigma(M, \alpha|_M; \mathrm{ord}(-n, \cdots, n)) \sigma(\overline{M}, \alpha|_{\overline{M}}; \mathrm{ord}(\overline{n}, \cdots, -\overline{n}))\\
& \times (-1)^{\int_{E_M + E_{\overline{M}}} \alpha} \times (c^{\dagger}_{-l})^{\alpha(e_{-l})} \cdots (c^{\dagger}_{l})^{\alpha(e_l)} \ket{0}
\bra{0} c_{l}^{\alpha(e_{\overline{l}})} \cdots c_{-l}^{\alpha(e_{-\overline{l}})},
\label{eq:reduced}
\end{aligned}
\end{equation}
where $N$ is given by gluing $M$ and $\overline{M}$ along the complement of $I$ on $\partial M$. Here, $\widetilde{Z}_N[\alpha]$ denotes the weight of the shadow theory evaluated on $N$.~\footnote{To be precise, $\widetilde{Z}_N[\alpha]$ in~\eqref{eq:reduced} should rather be written as $\widetilde{Z}_{M}[\alpha]\widetilde{Z}_{\overline{M}}[\alpha]$. However, we can redefine $\widetilde{Z}$ on the boundary to make~\eqref{eq:reduced} valid. Since $A=Cl(1)$ and $g^{ij}$ is diagonal $g^{ij}=\mathrm{diag}(g_i)$, we can do this by assigning an additional weight $\sqrt{g_i}$ on a boundary 1-simplex colored by $\epsilon_i\in A$. In the main text, we will work with such a redefinition.
}
$E_M, E_{\overline{M}}$ denotes a dual of $\eta$ introduced in~\eqref{eq:etadual}. 

\eqref{eq:reduced} reduces to the form of the path integral on $N$. To see this, first we associate the product of Grassmann integrals on $M$, $\overline{M}$ with that of $N$, by the following relation
\begin{equation}
\begin{aligned}
& \sigma(M, \alpha|_M; \mathrm{ord}(-n, \cdots, n)) \sigma(\overline{M}, \alpha|_{\overline{M}}; \mathrm{ord}(\overline{n}, \cdots, -\overline{n}))\\
 = & \prod_{l+1 \leq j \leq n}^{\mathrm{odd}} (-1)^{\alpha(e_j)} \prod_{l+1 \leq j \leq n}^{\mathrm{even}} (-1)^{\alpha(e_{-j})} \sigma(N, \alpha; \mathrm{ord}(-l, \cdots, l, \overline{l}, \cdots, -\overline{l})),
\end{aligned}
\label{eq:grassmannglue}
\end{equation}
which can be shown by an explicit computation of the Grassmann integral. Then,~\eqref{eq:reduced} is rewritten in the form of the path integral on $N$ (see Fig.~\ref{fig:MMN}),
\begin{equation}
\begin{aligned}
\rho_I = & \sum_{\alpha \in Z^1(N, \mathbb{Z}_2)} \widetilde{Z}_N[\alpha] \sigma(N, \alpha; \mathrm{ord}(-l, \cdots, l, \overline{l}, \cdots, -\overline{l})) (-1)^{\int_{E_N} \alpha}\\
& \times (c^{\dagger}_{-l})^{\alpha(e_{-l})} \cdots (c^{\dagger}_{l})^{\alpha(e_l)} \ket{0}
\bra{0} c_{l}^{\alpha(e_{\overline{l}})} \cdots c_{-l}^{\alpha(e_{-\overline{l}})},
\end{aligned}
\label{eq:reducedN}
\end{equation}
where we define $E_N$ as
\begin{align}
    E_N:=E_M+E_{\overline{M}}+\sum_{l+1 \leq j \leq n}^{\mathrm{odd}} e_j + \sum_{l+1 \leq j \leq n}^{\mathrm{even}} e_{-j}.
\end{align}
One can check that $\partial E_N$ correctly gives the dual of $w_2$ on $N$, when restricted to the interior of $N$. Thus, $E_N$ actually works as a dual of $\eta$ on $N$. 

\begin{figure}[htb]
\centering
\includegraphics{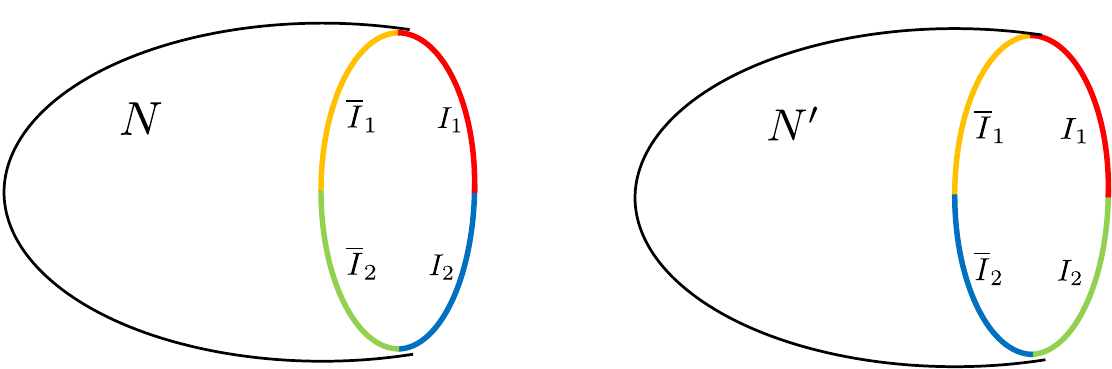}
\caption{Taking $\mathrm{tr}_{I}(\rho_{I} \rho_{I}^{\mathcal{T}_1})$ amounts to gluing $N$ and $N'$ along boundaries, such that the boundary 1-simplices with the same color in the figure are identified. We are gluing $I_1+\overline{I}_1$ (red and yellow curves) by the orientation reversing map, and $I_2+\overline{I}_2$ (blue and green curves) by the orientation preserving map. The resulting surface $X$ has a crosscap introduced along $I_1+\overline{I}_1$.}
\label{fig:glueNN}
\end{figure}

To compute the partial time-reversal of $\rho_I$, we express $\rho_I$ using the coherent state basis,
\begin{equation}
\begin{aligned}
\rho_I = & \sum_{\alpha \in Z^1(N, \mathbb{Z}_2)} \widetilde{Z}_N[\alpha] \sigma(N, \alpha; \mathrm{ord}(-l, \cdots, l, \overline{l}, \cdots, -\overline{l})) (-1)^{\int_{E_N} \alpha}\\
& \times \int \overleftarrow{d}\xi^{\alpha(e_{-l})}_{-l} \cdots \int \overleftarrow{d}\xi^{\alpha(e_{l})}_{l} \ket{\xi^{\alpha(e_{-l})}_{-l}}_{-l} \otimes \cdots \otimes \ket{\xi^{\alpha(e_{l})}_{l}}_{l}\\
& \times \int \overrightarrow{d}\overline{\xi}^{\alpha(e_{\overline{l}})}_{\overline{l}} \cdots \int \overrightarrow{d}\overline{\xi}^{\alpha(e_{-\overline{l}})}_{-\overline{l}} \subscripts{l}{\bra{\overline{\xi}^{\alpha(e_{\overline{l}})}_{\overline{l}}}}{} \otimes \cdots \otimes \subscripts{-l}{\bra{\overline{\xi}^{\alpha(e_{-\overline{l}})}_{-\overline{l}}}}{},
\end{aligned}
\end{equation}
where $\int \overleftarrow{d} \xi$ (resp.~$\int \overrightarrow{d} \overline{\xi}$) 
denotes the integral which satisfies $\xi \int \overleftarrow{d}\xi=1$ (resp.~$\int \overrightarrow{d}\overline{\xi}~ \overline{\xi} =1$).
Now we take the partial time-reversal~\eqref{eq:ptrans} acting on the region $I_1 = \sum_{1 \leq j \leq l} e_j$ and $\overline{I}_1 = \sum_{1 \leq j \leq l} e_{\overline{j}}$, 
{\footnotesize
\begin{equation}
\begin{aligned}
\rho_{I}^{\mathcal{T}_1} = & \sum_{\alpha \in Z^1(N, \mathbb{Z}_2)} \widetilde{Z}_N[\alpha] \sigma(N, \alpha; \mathrm{ord}(-l, \cdots, l, \overline{l}, \cdots, -\overline{l})) (-1)^{\int_{E_N} \alpha}\\
& \times \int \overleftarrow{d}\xi^{\alpha(e_{-l})}_{-l} \cdots \int \overleftarrow{d}\xi^{\alpha(e_{l})}_{l} \ket{\xi^{\alpha(e_{-l})}_{-l}}_{-l} \otimes \cdots \otimes \ket{\xi^{\alpha(e_{-1})}_{-1}}_{-1} \otimes \ket{i\overline{\xi}^{\alpha(e_{\overline{1}})}_{\overline{1}}}_{1} \otimes \cdots \otimes \ket{i\overline{\xi}^{\alpha(e_{\overline{l}})}_{\overline{l}}}_{l}\\
& \times \int \overrightarrow{d}\overline{\xi}^{\alpha(e_{\overline{l}})}_{\overline{l}} \cdots \int \overrightarrow{d}\overline{\xi}^{\alpha(e_{-\overline{l}})}_{-\overline{l}} {\subscripts{l}{\bra{i\xi^{\alpha(e_{l})}_{l}}}{}}\otimes \cdots \otimes {\subscripts{1}{\bra{i \xi^{\alpha(e_1)}_{1}}}{}} \otimes {\subscripts{-1}{\bra{\overline{\xi}^{\alpha(e_{-\overline{1}})}_{-\overline{1}}}}{}} \otimes \cdots \otimes {\subscripts{-l}{\bra{\overline{\xi}^{\alpha(e_{-\overline{l}})}_{-\overline{l}}}}{}}.
\end{aligned}
\end{equation}
}
which is rewritten in the fermion number basis as
\begin{equation}
\begin{aligned}
\rho_{I}^{\mathcal{T}_1} = & \sum_{\alpha \in Z^1(N, \mathbb{Z}_2)} \widetilde{Z}_N[\alpha] \sigma(N, \alpha; \mathrm{ord}(-l, \cdots, -1, \overline{1}, \cdots, \overline{l},l,\cdots, 1,-\overline{1}, \cdots, -\overline{l}))\\
& \times  (-1)^{\int_{E_N} \alpha} \prod_{e\in I_1\cup\overline{I}_1} (-i)^{\alpha(e)}  \times (c^{\dagger}_{-l})^{\alpha(e_{-l})} \cdots (c^{\dagger}_{-1})^{\alpha(e_{-1})} (c^{\dagger}_{1})^{\alpha(e_{\overline{1}})} \cdots (c^{\dagger}_{l})^{\alpha(e_{\overline{l}})} \ket{0}\\
& \times \bra{0} c_{l}^{\alpha(e_{l})} \cdots c_{1}^{\alpha(e_{1})} c_{-1}^{\alpha(e_{-\overline{1}})} \cdots c_{-l}^{\alpha(e_{-\overline{l}})}.
\end{aligned}
\label{eq:ptrfermi}
\end{equation}
Now we can explicitly write down the 
order parameter~\eqref{eq:ptr} as 
\begin{equation}
\begin{aligned}
\mathrm{tr}_{I}(\rho_{I} \rho_{I}^{\mathcal{T}_1}) = 
& \sum_{\alpha \in Z^1(N, \mathbb{Z}_2)} \sum_{\alpha' \in Z^1(N', \mathbb{Z}_2)} \widetilde{Z}_{N}[\alpha] \widetilde{Z}_{N'}[\alpha'] \sigma(N, \alpha; \mathrm{ord}(-l, \cdots, l, \overline{l}, \cdots, -\overline{l}))\\
& \times \sigma(N', \alpha'; \mathrm{ord}(-l, \cdots, -1, \overline{1}, \cdots, \overline{l},l,\cdots, 1,-\overline{1}, \cdots, -\overline{l}))\\
& \times(-1)^{\int_{E_N}\alpha}(-1)^{\int_{E_{N'}}\alpha'} \prod_{e\in I_1\cup\overline{I}_1} (-i)^{\alpha(e)} \\
& \times \prod_{1 \leq j \leq l} \delta_{\alpha(e_{-j}) \alpha'(e_{-\overline{j}})} \delta_{\alpha(e_{j}) \alpha'(e_{j})} \delta_{\alpha(e_{-\overline{j}}) \alpha'(e_{-j})} \delta_{\alpha(e_{\overline{j}}) \alpha'(e_{\overline{j}})},
\label{eq:ptrtft}
\end{aligned}
\end{equation}
where the expression involves two copies of $N$ evaluating $\rho_{I}^{\mathcal{T}_1}$ and $\rho_I$ written as $N$ and $N'$ respectively.
By taking the trace after partial time-reversal, the expression~\eqref{eq:ptrtft} looks like the form of path integral on a space $X$, which is obtained by gluing $N, N'$ along their boundaries as illustrated in Fig.~\ref{fig:glueNN}.
Namely, we identify $I_1+ \overline{I}_1=\sum_{1 \leq j \leq l} (e_j + e_{\overline{j}})$ on $\partial N$ and $\partial N'$,  by the orientation reversing map, and $I_2+\overline{I}_2=\sum_{1 \leq j \leq l} (e_{-j} + e_{-\overline{j}})$ by the orientation preserving map.
Here, the induced map $N\sqcup N'\to X$ is restricted to each $N$ as $N\to \widetilde{N}$, where $\widetilde{N}$ is given by identifying two boundary 0-simplices of $N$ contained in $\partial (I_1+\overline{I}_1)$. Then,~\eqref{eq:ptrtft} is rewritten in the form of path integral on $X$ as
\begin{equation}
\begin{aligned}
\mathrm{tr}_{I}(\rho_{I} \rho_{I}^{\mathcal{T}_1}) = 
& \sum_{\alpha \in Z^1(X, \mathbb{Z}_2)} \widetilde{Z}_X[\alpha]  \sigma(\widetilde{N}, \alpha|_{\widetilde{N}}; \mathrm{ord}(-l, \cdots, l, \overline{l}, \cdots, -\overline{l}))\\
& \times \sigma(\widetilde{N}', \alpha|_{\widetilde{N}'}; \mathrm{ord}(-l, \cdots, -1, \overline{1}, \cdots, \overline{l},l,\cdots, 1,-\overline{1}, \cdots, -\overline{l}))\\
& \times(-1)^{\int_{E_{\widetilde{N}}}\alpha}(-1)^{\int_{E_{\widetilde{N}'}}\alpha}\times \prod_{e\in I_1\cup\overline{I}_1} (-i)^{\alpha(e)}.
\label{eq:ptrlikepintft}
\end{aligned}
\end{equation}

The expression~\eqref{eq:ptrlikepintft} looks like the partition function of pin TQFT constructed from the Grassmann integral~\eqref{sigmadefpin}, since both assigns $\pm i$ factor to 1-simplices where we reverse the orientation. We can actually show that these are identical. To see this, we compare~\eqref{eq:ptrlikepintft} with the pin$^-$ TQFT on $X$. 
By integrating out the Grassmann variables living on boundaries of $\widetilde{N}$ and $\widetilde{N}'$, the pin$^-$ TQFT partition function is obtained as
\if
which is given by
{\small
\begin{equation}
\begin{aligned}
    Z[X, \eta]=
    & \sum_{\alpha \in Z^1(X, \mathbb{Z}_2)} \widetilde{Z}_{X}[\alpha] \sigma(\widetilde{N}, \alpha|_{\widetilde{N}}; \mathrm{ord}(-l, \cdots, l, \overline{l}, \cdots, -\overline{l}))\\
    & \times \sigma(\widetilde{N}', \alpha|_{\widetilde{N}'}; \mathrm{ord}(-l, \cdots, l, \overline{l}, \cdots, -\overline{l})) (-1)^{\int_{E_{X}}\alpha}\\
& \times \prod_{e_j\in I_1}^{\mathrm{odd}}(+i)^{\alpha(e_j)}\prod_{e_j\in I_1}^{\mathrm{even}}(-i)^{\alpha(e_j)}
\prod_{e_{\overline{j}}\in \overline{I}_1}^{\mathrm{odd}}(-i)^{\alpha(e_{\overline{j}})}\prod_{e_{\overline{j}}\in \overline{I}_1}^{\mathrm{even}}(+i)^{\alpha(e_{\overline{j}})}\\
&\times \prod_{\partial \widetilde{N}\cup\partial \widetilde{N}'}d\theta d\overline{\theta}\cdot(\theta_{-l}^{\alpha(e_{-l})}\cdots\overline{\theta}_{l}^{\alpha(e_{l})})(\theta_{\overline{l}}^{\alpha(e_{\overline{l}})}\cdots\overline{\theta}_{-\overline{l}}^{\alpha(e_{-\overline{l}})})\\
&\times (\theta_{-\overline{l}}^{\alpha(e_{-\overline{l}})}\dots\overline{\theta}_{-\overline{1}}^{\alpha(e_{-\overline{1}})}\overline{\theta}_1^{\alpha(e_1)}\cdots \theta_l^{\alpha(e_l)})(\overline{\theta}_{\overline{l}}^{\alpha(e_{\overline{l}})}\cdots\theta_{\overline{1}}^{\alpha(e_{\overline{1}})}{\theta_{-1}^{\alpha(e_{-1})}}\cdots\overline{\theta}_{-l}^{\alpha(e_{-l})}),
\end{aligned}
\end{equation}
}
where $E_X$ is the dual of $w_2+w_1^2$ on $X$, whose choice of representative is described in Sec.~\ref{sec:tqft}. By performing the Grassmann integral on boundaries, we obtain
\fi
\begin{equation}
\begin{aligned}
    Z[X,\eta]=
    & \sum_{\alpha \in Z^1(X, \mathbb{Z}_2)} \widetilde{Z}_{X}[\alpha] \sigma(\widetilde{N}, \alpha|_{\widetilde{N}}; \mathrm{ord}(-l, \cdots, l, \overline{l}, \cdots, -\overline{l}))\\
    & \times \sigma(\widetilde{N}', \alpha|_{\widetilde{N}'}; \mathrm{ord}(-l, \cdots, -1, \overline{1}, \cdots, \overline{l},l,\cdots, 1,-\overline{1}, \cdots, -\overline{l})) \\
&  \times (-1)^{\int_{E_{X}}\alpha}\prod_{e\in I_2}^{\mathrm{even}}(-1)^{\alpha(e)}\prod_{e\in \overline{I}_2}^{\mathrm{odd}}(-1)^{\alpha(e)}\times \prod_{e\in I_1\cup\overline{I}_1} (-i)^{\alpha(e)},
\label{eq:pintftlikeptr}
\end{aligned}
\end{equation}
where $E_X$ is the dual of $\eta$ trivializing $w_2+w_1^2$ on $X$, whose choice of representative is described in Sec.~\ref{sec:tqft}. 
By comparing~\eqref{eq:ptrlikepintft} with~\eqref{eq:pintftlikeptr}, one can see these expression are completely the same, by checking that
\begin{align}
    E_{\widetilde{N}}+E_{\widetilde{N}'}+\sum_{e_j\in I_2}^{\mathrm{even}}e_j+\sum_{e_j\in \overline{I}_2}^{\mathrm{odd}}e_j=E_X.
\end{align}
Thus, we have shown that~\eqref{eq:ptr} is identical to the partition function of the pin$^-$ TQFT,
\begin{align}
    \mathrm{tr}_{I}(\rho_{I} \rho_{I}^{\mathcal{T}_1}) = Z[X,\eta].
\end{align}
For instance, we can evaluate $\mathrm{tr}_{I}(\rho_{I} \rho_{I}^{\mathcal{T}_1})$ for the ground state wave function
of the Kitaev chain described in Sec.~\ref{subsec:kitaev}. Using the form of the ABK invariant~\eqref{eq:abk}, the expression becomes
\begin{equation}
\mathrm{tr}_{I} (\rho_I \rho_{I}^{\mathcal{T}_1}) = 2^{-\chi(Y) + \chi(X) / 2} \mathrm{ABK}[X, \eta],
\end{equation}
where $Y$ is a closed surface given by gluing $M$ and $\overline{M}$ along the boundaries.
In particular, if we choose $M$ as a disk, $Y = S^2$ and $X = \mathbb{RP}^2$, which gives 
\begin{equation}
\mathrm{tr}_{I} (\rho_I \rho_{I}^{\mathcal{T}_1}) = \frac{1}{2\sqrt{2}} e^{\pm 2\pi i / 8}.
\end{equation}
This reproduces the results obtained in \cite{Shapourian-Shiozaki-Ryu2017a}.

\section{Entanglement negativity}
\label{sec:nega}

In this section, we evaluate the R\'enyi entanglement negativity of even degree $n$, which is defined as
\begin{equation}
\mathcal{E}_{n} = \log \mathrm{tr} [\rho_I^{\mathcal{T}_1} (\rho_I^{\mathcal{T}_1})^{\dagger} \cdots \rho_I^{\mathcal{T}_1} (\rho_I^{\mathcal{T}_1})^{\dagger}],
\end{equation}
where $\rho_I^{\mathcal{T}_1} (\rho_I^{\mathcal{T}_1})^{\dagger}$ is multiplied $n/2$ times in the above expression.
Let us comment on notations used throughout this section.
\begin{itemize}
\item
Following the notation in Sec.~\ref{subsec:ptrcomp},
$\rho_I^{\mathcal{T}_1}$ and $(\rho_I^{\mathcal{T}_1})^{\dagger}$ are thought to be a path integral on surfaces $N$ and $N^{\ast}$ respectively, where $N^{\ast}$ is given by reversing the orientation of $N$. For simplicity, $N$ is taken to be a disk. 
We represent the boundary intervals as $I_1, \overline{I}_1, I_2, \overline{I}_2\in \partial N$ (following Fig~\ref{fig:MMN} (b)), and $I_{1^*}, \overline{I}_{1^*}, I_{2^*}, \overline{I}_{2^*}\in \partial N^*$.
\item 
We introduce the following notation for path integral on an open surface 
\begin{align}
    z[M,\eta_M,\alpha; \mathrm{ord}(e_1,e_2,\dots, e_n)]:=\sigma(M,\alpha;\mathrm{ord}(e_1,e_2,\dots, e_n))(-1)^{\int_{E_M}\alpha},
\end{align}
where $e_1,e_2,\dots, e_n$ are boundary 1-simplices of $M$.
It is convenient to mention the behavior of the path integral under gluing surfaces. Let us consider gluing two open surfaces $M$, $M'$ along the boundary interval $I$, whose 1-simplices are denoted as $\tilde{e}_1,\tilde{e}_2,\dots, \tilde{e}_m$. If we denote the resulting surface $X$, we have the following relation between the path integral before and after gluing,
\begin{equation}
\begin{aligned}
   z[M,\eta_M,\alpha; \mathrm{ord}(e_1,\dots, e_l,\tilde{e}_1,\dots, \tilde{e}_m)]&
   z[M',\eta_{M'},\alpha; \mathrm{ord}(\tilde{e}_m,\dots, \tilde{e}_1,e_{1'},\dots, e_{l'})]\\
    & = z[X,\eta_X,\alpha; \mathrm{ord}(e_1,\dots, e_l,e_{1'},\dots, e_{l'})],
    \label{eq:glueintegral}
\end{aligned}
\end{equation}
where we \textit{defined} $E_X$ as
\begin{align}
    E_X:=E_M+E_{M'}+\sum_{\tilde{e}\in I}^{+}\tilde{e},
\end{align}
where the sum runs over boundary 1-simplices of $M$ contained in $I$, rounding a 2-simplex of $M$ whose sign is $+$.
\end{itemize}
We aim to show that the quantity $e^{\mathcal{E}_{n}}$ is identified as a path integral of the TQFT on a certain closed surface. To do this, we start with examining how the surface looks like.
We can obtain the resulting surface by gluing each $N$ and $N^*$ step by step; (i) multiplying $\rho_I^{\mathcal{T}_1}$ and $(\rho_I^{\mathcal{T}_1})^{\dagger}$, (ii) multiplying $\rho_I^{\mathcal{T}_1}(\rho_I^{\mathcal{T}_1})^{\dagger}$s and taking the trace.

\begin{figure}[htb]
\centering
\includegraphics{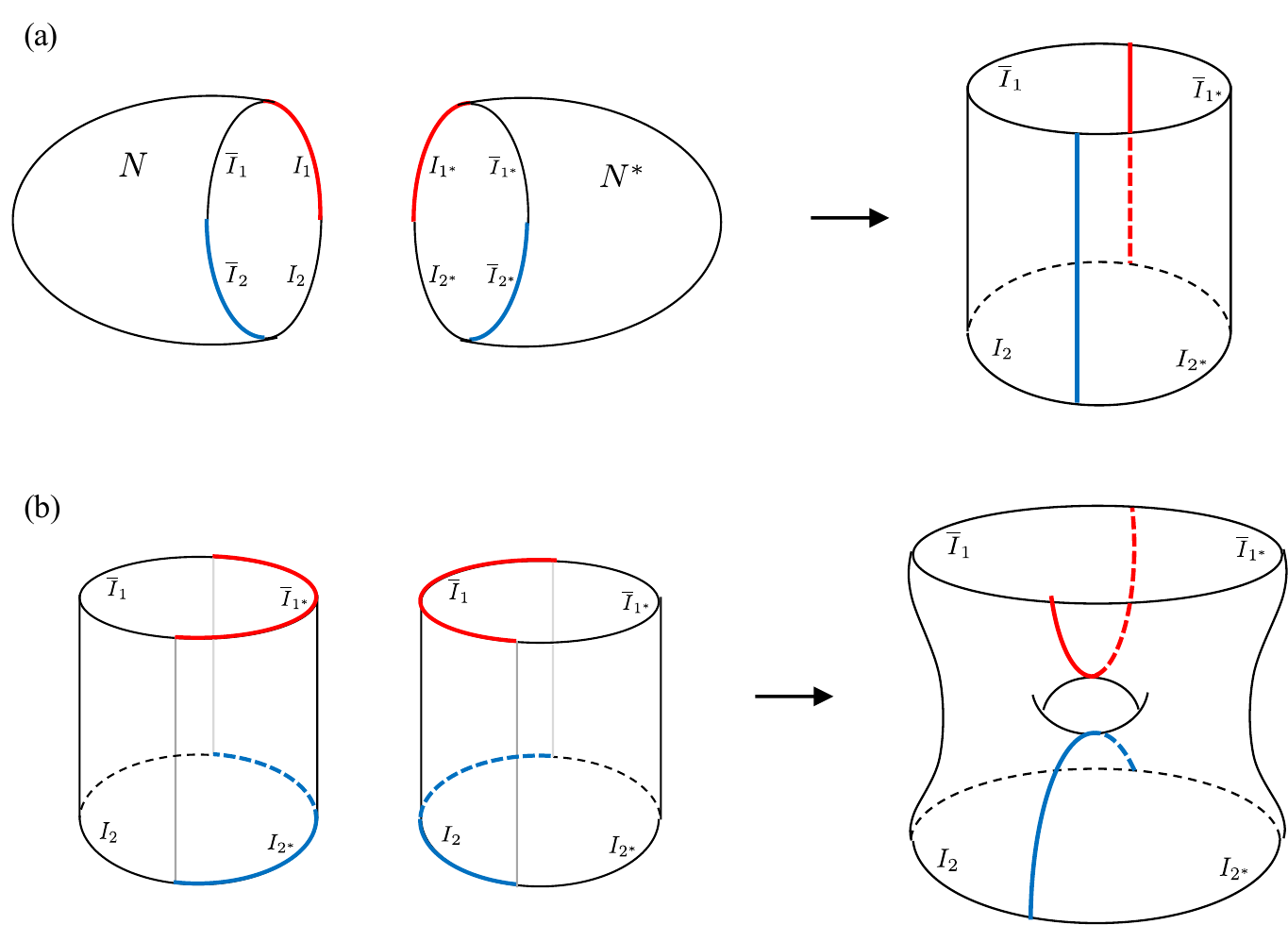}
\caption{(a): 
The computation of $\rho_I^{\mathcal{T}_1}(\rho_I^{\mathcal{T}_1})^\dagger$ gives a path integral on a cylinder. (b): Taking $\rho_I^{\mathcal{T}_1}(\rho_I^{\mathcal{T}_1})^\dagger\rho_I^{\mathcal{T}_1}(\rho_I^{\mathcal{T}_1})^\dagger$ gives a path integral on a torus with two punctures.}
\label{fig:gluenegativity}
\end{figure}

Let us begin with the first step.
Since $\rho_I^{\mathcal{T}_1}$~\eqref{eq:ptrfermi} has an outgoing state in the interval $I_1$ and $\overline{I}_2$ of $\partial N$, taking $\rho_I^{\mathcal{T}_1}(\rho_I^{\mathcal{T}_1})^\dagger$ amounts to gluing $N$ and $N^*$ along $I_1, \overline{I}_2$  and $I_{1^*}, \overline{I}_{2^*}$, making up a cylinder. See Fig.~\ref{fig:gluenegativity} (a).
Similarly, one finds that gluing two $\rho_I^{\mathcal{T}_1}(\rho_I^{\mathcal{T}_1})^{\dagger}$s gives a torus with two punctures, by gluing two copies of cylinder along $\overline{I}_{1^*}, I_{2^*}$ and $\overline{I}_{1}, I_{2}$, see Fig.~\ref{fig:gluenegativity} (b).
Then, $\mathcal{E}_{n}$ amounts to successive gluing of $n/2$ cylinders, which gives an oriented closed surface $\Sigma_{g(n)}$ with genus $g(n)=n/2-1$.
Actually, we will show that the R\'enyi negativity is directly associated with the partition function of spin TQFT as
\begin{equation}
e^{\mathcal{E}_{n}} = Z[\Sigma_{g(n)}, \eta]=\sum_{\alpha \in Z^1(\Sigma_{g(n)}, \mathbb{Z}_2)} \widetilde{Z}_{\Sigma_{g(n)}}[\alpha]z[\Sigma_{g(n)},\eta,\alpha],
\label{eq: moment}
\end{equation}
which will be demonstrated in the following subsection. Though we will mainly work on the Kitaev chain wave function, the above correspondence is safely extended for generic spin TQFT prepared by a $\mathbb{Z}_2$-graded Frobenius algebra (see~\ref{app:ptrfrob}).

\subsection{Evaluation of entanglement negativity}
Let us turn to the explicit computation of $e^{\mathcal{E}_n}$.
For notational simplicity, we ignore the shadow theory part in
$\rho_I^{\mathcal{T}_1}$, $(\rho_I^{\mathcal{T}_1})^{\dagger}$ shown in~\eqref{eq:ptrfermi} and just focus on the Grassmann integral part, and write
\begin{equation}
\begin{aligned}
\rho_I^{\mathcal{T}_1} & \approx z[N, \eta_N, \alpha; \mathrm{ord}(-l, \cdots, -1, \overline{1}, \cdots, \overline{l},l,\cdots, 1,-\overline{1}, \cdots, -\overline{l})],\\
(\rho_I^{\mathcal{T}_1})^{\dagger} & \approx z[N^*, \eta_{N^*}, \alpha; \mathrm{ord}(-\overline{l}^*, \cdots, -\overline{1}^*, 1^*, \cdots, l^*,\overline{l}^*,\cdots, \overline{1}^*,-1^*, \cdots, -l^*)].
\end{aligned}
\label{eq:reducedgrassmann}
\end{equation}
Here, we omitted the imaginary factor 
$\prod_{e\in I_1\cup \overline{I}_1} (-i)^{\alpha(e)}$ 
in $\rho_I^{\mathcal{T}_1}$ 
in~\eqref{eq:ptrfermi}, since this contribution cancels out with that of $(\rho_I^{\mathcal{T}_1})^{\dagger}$ when evaluating the moment.
From now on, let us simply write the rhs of~\eqref{eq:reducedgrassmann} as $z[N,\eta_N,\alpha]$ and $z[N^*,\eta_{N^*},\alpha]$ respectively. If we write the cylinder as $X_0$ obtained by gluing $N$ and $N^*$ as shown in Fig.~\ref{fig:gluenegativity}, using~\eqref{eq:glueintegral} we have
\begin{align}
    z[N,\eta_N, \alpha]z[N^*,\eta_{N^*}, \alpha]=z[X_0,\eta_{X_0}, \alpha],
\end{align}
where the ordering of boundary 1-simplices of $X_0$ is induced from those of $N$, $N^*$. This allows us to write $\rho_I^{\mathcal{T}_1}(\rho_I^{\mathcal{T}_1})^{\dagger}$ in a similar fashion to~\eqref{eq:reducedgrassmann} as
\begin{align}
    \rho_I^{\mathcal{T}_1}(\rho_I^{\mathcal{T}_1})^{\dagger}\approx z[X_0,\eta_{X_0}, \alpha]. 
\end{align}
\if
For later convenience, we divide $\mathrm{sgn}[\alpha(e_1) \cdots \alpha(e_{\overline{1}})]$ into three terms:
\begin{equation}
\mathrm{sgn}[\alpha(e_1) \cdots \alpha(e_{\overline{1}})] = \mathrm{sgn}[\alpha(e_1) \cdots \alpha(e_l)] \mathrm{sgn}[\alpha(e_{\overline{l}}) \cdots \alpha(e_{\overline{1}})] \mathrm{sgn}[\{ \alpha(e_j) \} \{ \alpha(e_{\overline{j}}) \}],
\end{equation}
where $\mathrm{sgn}[\{ \alpha(e_j) \} \{ \alpha(e_{\overline{j}}) \}]$ denotes the fermionic sign arising from reordering 
\begin{align}
    (\epsilon_1\cdots\epsilon_l)(\epsilon_{\overline{l}}\cdots\epsilon_{\overline{1}})\to (\epsilon_{\overline{l}}\cdots\epsilon_{\overline{1}})(\epsilon_1\cdots\epsilon_l), 
\end{align}
where $\epsilon_j$ carries the fermion parity $\alpha(e_j)$.
Then, one can express the product of $\rho_I^{R_1}$ and $(\rho_I^{R_1})^{\dagger}$ as 
\begin{equation}
\begin{aligned}
\rho_I^{R_1} (\rho_I^{R_1})^{\dagger} \approx & \sigma(N, \alpha; \mathrm{ord}(-l, \cdots, -\overline{l})) \sigma(N^*, \alpha; \mathrm{ord}(-\overline{l}^*, \cdots, -l^*)) (-1)^{\int_{E_N} \alpha} (-1)^{\int_{E_{N^*}} \alpha}\\
& \times \mathrm{sgn}[\alpha(e_{1}) \cdots \alpha(e_{l})] \mathrm{sgn}[\alpha(e_{{\overline{l}}}) \cdots \alpha(e_{{\overline{1}}})] \mathrm{sgn}[\{ \alpha(e_{j}) \} \{ \alpha(e_{{\overline{j}}}) \}]\\
& \times \mathrm{sgn}[\alpha(e_{1^*}) \cdots \alpha(e_{l^*})] \mathrm{sgn}[\alpha(e_{{\overline{l}}^*}) \cdots \alpha(e_{{\overline{1}}^*})] \mathrm{sgn}[\{ \alpha(e_{j^*}) \} \{ \alpha(e_{{\overline{j}}^*}) \}].
\end{aligned}
\end{equation}
Considering that $e_{j}$ and $e_{j^*}$ ($1 \leq j \leq l$) are glued together, one can see that the sign factors $\mathrm{sgn}[\alpha(e_{1}) \cdots \alpha(e_{l})]$ and $\mathrm{sgn}[\alpha(e_{1^*}) \cdots \alpha(e_{l^*})]$ cancel out. 
Moreover, one can rewrite the remaining part as the product of the Grassmann integral and the spin structure factor on $X_0$, which is a cylinder obtained by gluing $N$ and $\overline{N}$ (see Fig.~\ref{fig:gluenegativity} (a)),
\begin{equation}
\begin{aligned}
& \sigma(N, \alpha; \mathrm{ord}(-l, \cdots, -\overline{l})) \sigma(N^*, \alpha; \mathrm{ord}(-\overline{l}^*, \cdots, -l^*)) (-1)^{\int_{E_N} \alpha} (-1)^{\int_{E_{N^*}} \alpha}\\
&\times  \mathrm{sgn}[\{ \alpha(e_{j}) \} \{ \alpha(e_{{\overline{j}}}) \}] \mathrm{sgn}[\{ \alpha(e_{j^*}) \} \{ \alpha(e_{{\overline{j}}^*}) \}]\\
= & \sigma(X_0, \alpha; \mathrm{ord}(-l, \cdots, -1, \overline{l}, \cdots, \overline{1}, \overline{1}^*, \cdots, \overline{l}^*, -1^*, \cdots, -l^*)) (-1)^{\int_{E_{X_0}} \alpha}.
\end{aligned}
\end{equation}
This computation is performed essentially the same fashion as the gluing of $M$ and $\overline{M}$ in the analysis of partial time-reversal~\eqref{eq:grassmannglue}.
Now we find
\begin{equation}
\begin{aligned}
\rho_I^{R_1} (\rho_I^{R_1})^{\dagger} \approx & \sigma(X_0, \alpha; \mathrm{ord}(-l, \cdots, -1, \overline{l}, \cdots, \overline{1}, \overline{1}^*, \cdots, \overline{l}^*, -1^*, \cdots, -l^*))\\
& \times (-1)^{\int_{E_{X_0}} \alpha} \mathrm{sgn}[\alpha(e_{{\overline{l}}}) \cdots \alpha(e_{{\overline{1}}})] \mathrm{sgn}[\alpha(e_{{\overline{l}}^*}) \cdots \alpha(e_{{\overline{1}}^*})].
\end{aligned}
\end{equation}
If we define $z[X_0, \alpha, \eta]$ as the product of the Grassmann integral and the spin structure factor, we can simply express 
\begin{equation}
\rho_I^{R_1} (\rho_I^{R_1})^{\dagger} \approx z[X_0, \alpha, \eta]\cdot \mathrm{sgn}[\alpha(e_{{\overline{l}}}) \cdots \alpha(e_{{\overline{1}}})] \mathrm{sgn}[\alpha(e_{{\overline{l}}^*}) \cdots \alpha(e_{{\overline{1}}^*})].
\end{equation}
\fi
If we further glue two cylinders like Fig.~\ref{fig:gluenegativity} (b), we have
\begin{equation}
z[X_0, \eta_{X_0}, \alpha] z[X^{\prime}_0, \eta_{X'_0}, \alpha] = z[X_1, \eta_{X_1}, \alpha],
\end{equation}
where $X_0$, $X'_0$ denotes two copies of cylinder, and $X_1$ is a two-punctured torus given by gluing two cylinders. By successive gluing of cylinders, if we let $X_{g}$ denote a genus $g$ surface with two punctures, one finds 
\begin{equation}
\rho_I^{\mathcal{T}_1}(\rho_I^{\mathcal{T}_1})^{\dagger} \cdots \rho_I^{\mathcal{T}_1}(\rho_I^{\mathcal{T}_1})^{\dagger} \approx z[X_{g(n)}, \eta_{X_{g(n)}}, \alpha].
\label{eq: procedure2}
\end{equation}
Finally, we obtain a closed surface with genus $g(n) = n / 2 - 1$ by taking the trace of the equation (\ref{eq: procedure2}),
\begin{equation}
\mathrm{tr} [ \rho_I^{\mathcal{T}_1}(\rho_I^{\mathcal{T}_1})^{\dagger} \cdots \rho_I^{\mathcal{T}_1}(\rho_I^{\mathcal{T}_1})^{\dagger} ] \approx z[\Sigma_{g(n)}, \eta_{\Sigma_{g(n)}}, \alpha].
\label{eq: procedure3}
\end{equation}
Now let us examine what the induced structure $\eta_{\Sigma_{g(n)}}$ is like. First, it is not hard to see that $\eta_{\Sigma_{g(n)}}$ correctly gives a trivialization of $w_2$, $\delta\eta=w_2$, thereby defining a spin structure on $\Sigma_{g(n)}$. We can further show that the induced spin structure $\eta$ is equivalent to a connected sum of $g(n)$ copies of (NS, NS) torus. To see this, let us first determine the spin structure measured around a cylinder $X_0$ given by gluing $N$ and $N^*$ (Fig.~\ref{fig:gluenegativity} (a)). In this case, if we denote $C_{\chi}$ as a cycle around a cylinder, we have
\begin{align}
    z[N,\eta_N,\chi]z[N^*,\eta_{N^*},\chi]=z[X_0,\eta_{X_0},\chi],
    \label{eq:determineNS}
\end{align}
where $\chi$ is a dual 1-cocycle of $C_{\chi}$. Since the configuration of $C_{\chi}$ can be chosen in a symmetric fashion such that $z[N,\eta_N,\chi]$ and $z[N^*,\eta_{N^*},\chi]$ are identical, we see that the lhs of~\eqref{eq:determineNS} is 1. Thus, we have $z[X_0,\eta_{X_0},\chi]=1$, which means that the induced spin structure is NS around $C_{\chi}$. Using the same logic for the other cycles of $\Sigma_{g(n)}$, we see that the spin structure is NS for all fundamental cycles of $\Sigma_{g(n)}$, hence we have a connected sum of $g(n)$ copies of (NS, NS) tori.

Recalling that a bosonic shadow theory and the summation over $\alpha$ is omitted in the equation (\ref{eq: procedure3}), we obtain 
\begin{equation}
\mathrm{tr} [ \rho_I^{\mathcal{T}_1}(\rho_I^{\mathcal{T}_1})^{\dagger} \cdots \rho_I^{\mathcal{T}_1}(\rho_I^{\mathcal{T}_1})^{\dagger} ] = \sum_{\alpha \in Z^1(\Sigma_{g(n)}, \mathbb{Z}_2)} \widetilde{Z}_{\Sigma_{g(n)}} [\alpha] z[\Sigma_{g(n)}, \eta, \alpha] = Z [\Sigma_{g(n)}, \eta].
\end{equation} 
This is what we wanted to achieve.
We have shown that the moments of partial time-reversal are identical to the partition functions on a surface with genus $g(n)$ with NS spin structure for all fundamental cycles. 
If we employ the ground state wave function of 
the Kitaev chain described in Sec.~\ref{subsec:kitaev}, we obtain
\begin{equation}
e^{\mathcal{E}_{n}} = \sqrt{2}^{4-3n}.
\end{equation}

\section{Moments of partial time-reversal}
\label{sec:moment}
In this section, we compute the moments of partial time-reversal for any power
\begin{equation}
\begin{aligned}
Z_{n} = \mathrm{tr} [\rho_I^{\mathcal{T}_1} \cdots \rho_I^{\mathcal{T}_1}],\\
\widetilde{Z}_{n} = \mathrm{tr} [\rho_I^{\widetilde{\mathcal{T}}_1}\cdots \rho_I^{\widetilde{\mathcal{T}}_1}],
 \end{aligned}
\end{equation}
where $\rho_I^{\widetilde{\mathcal{T}}_1}$ denotes the partially transposed density matrix twisted by the fermion number parity $(-1)^{F_1}$ of the interval $I_1$~\cite{Shapourian-Ruggiero-Ryu-Calabrese2019}, 
\begin{equation}
\rho_I^{\widetilde{\mathcal{T}}_1} := \rho_I^{\mathcal{T}_1} (-1)^{F_1}.
\end{equation}
Here, we note that $\widetilde{Z}_{n}$ coincides with the moment of partial time-reversal $e^{\mathcal{E}_{n}}$ computed in Sec.~\ref{sec:nega} when $n$ is even, because 
\begin{equation}
(\rho_I^{\mathcal{T}_1})^{\dagger} = (-1)^{F_1} \rho_I^{\mathcal{T}_1} (-1)^{F_1}.
\end{equation}
As well as the case of $\mathcal{E}_{n}$, these moments can also be represented as a partition function on a surface $\Sigma_g$ with genus $g$ where $g = n/2 -1$ for an even $n$, and $g = (n-1)/2$ for an odd $n$.
Interestingly, we find that $Z_n$ shows the following $\mathbb{Z}_8$ effect, while $\widetilde{Z}_n$ just shows even/odd effect,
\begin{equation}
\begin{aligned}
Z_n = (2\sqrt{2})^{-2g} \times 
\begin{cases}
+ 1 / 2 \quad &n = 0 \bmod 8\\
+ 1 \quad &n = \pm 1 \bmod 8\\
0 \quad &n = \pm 2 \bmod 8\\
- 1 \quad &n = \pm 3 \bmod 8\\
- 1 / 2 \quad &n = 4 \bmod 8
\end{cases}
\end{aligned}
\label{eq:momentperiod8}
\end{equation}
\begin{equation}
\begin{aligned}
\widetilde{Z}_n = (2\sqrt{2})^{-2g} \times 
\begin{cases}
1 / 2 \quad &n = 0 \bmod 2\\
0 \quad &n = 1 \bmod 2,
\end{cases}
\end{aligned}
\label{eq:momentperiod2}
\end{equation}
for the wave function of Kitaev chain described in Sec.~\ref{subsec:kitaev}.

Here, let us outline the key steps of the computation, focusing on $Z_{n}$.
The evaluation of $Z_n$ runs largely parallel to the case of 
the entanglement negativity $e^{\mathcal{E}_n}$ in Sec.~\ref{sec:nega}. 
First, let us drop
the imaginary factor $\prod_{e\in I_1\cup \overline{I}_1} (-i)^{\alpha(e)}$ in $\rho_I^{\mathcal{T}_1}$~\eqref{eq:ptrfermi} supported on $I_1\cup \overline{I}_1$. Then, analogously to Sec.~\ref{sec:nega}, $Z_n$ is regarded as the path integral on some surface $X$ given by gluing $n$ copies of $N$. During the process of the successive gluing of surfaces, let us denote the intermediate surface obtained by gluing $k$ copies of $N$ which corresponds to $(\rho_I^{\mathcal{T}_1})^k$ as $N^{(k)}$. Then, every time we glue 
the $(k+1)$th copy of $N$ with $N^{(k)}$ along the interval in $\partial N^{(k)}$, to evaluate $(\rho_I^{\mathcal{T}_1})^{k+1}$, we have the following relation between the partition function before and after gluing using~\eqref{eq:glueintegral},
\begin{align}
    z[N^{(k)},\eta_{N^{(k)}},\alpha]z[N,\eta_N,\alpha]=z[N^{(k+1)},\eta_{N^{(k+1)}},\alpha].
\end{align}
By successively applying this relation and taking the trace in the last step, we finally obtain $E_{X}$, dual of $\eta_X$ induced on the resulting surface $X$ which corresponds to $Z_n$. Hence,
if we ignore the imaginary factor $\prod_{e\in I_1\cup \overline{I}_1} (-i)^{\alpha(e)}$ in $\rho_I^{\mathcal{T}_1}$~\eqref{eq:ptrfermi}, 
$Z_{n}$ would be associated with the partition function on $X$ as
\begin{align}
    Z[X,\eta_X]=\sum_{\alpha\in Z^1(X,\mathbb{Z}_2)} \widetilde{Z}_X[\alpha] z[X,\eta_X,\alpha]=\sum_{\alpha\in Z^1(X,\mathbb{Z}_2)} \widetilde{Z}_X[\alpha] \sigma(X,\alpha)(-1)^{\int_{E_X}\alpha}.
    \label{eq:naiveZn}
\end{align}
The above expression does not necessarily give the partition function of spin TQFT, since $E_{X}$ constructed above may or may not provide the correct trivialization of $w_2$. 

Next, we incorporate the effect of the imaginary factor $\prod_{e\in I_1\cup \overline{I}_1} (-i)^{\alpha(e)}$ in $\rho_I^{\mathcal{T}_1}$~\eqref{eq:ptrfermi} supported on $I_1\cup \overline{I}_1$. The imaginary factor introduces the shift of the spin structure. This factor acts as the ``half of fermion parity'' on the fermions living in $ I_1\cup \overline{I}_1$. Hence, if we glue two copies of $N$ along $I_1$ and $\overline{I}_1$, the doubled imaginary factor gives the fermion parity twist to the resulting surface, inserted in the interval where we glued them, see Fig.~\ref{fig:glueRR} (a).
Accordingly, every time we glue surfaces along $I_1\subset \partial N^{(k)}$ and $\overline{I}_1\subset\partial N$, we introduce the twist by fermion parity in the resulting surface. 
The fermion parity twist leads to the shift of $\eta_X$, which is expressed as 
$\eta_X\to\eta_X+\chi$ where $\chi$ is the dual of the 1-cycle $C_{\chi}$ in which we insert the twist.
Incorporating this effect to the expression~\eqref{eq:naiveZn}, we find that $Z_n$ is given by
\begin{align}
    Z_n=
    Z[X,\eta_X+\chi]=\sum_{\alpha\in Z^1(X,\mathbb{Z}_2)} \widetilde{Z}_X[\alpha] \sigma(X,\alpha)(-1)^{\int_{E_X+C_{\chi}}\alpha}.
    \label{eq:correctZn}
\end{align}
The above expression becomes the partition function of spin TQFT, if $E_X+C_{\chi}$ gives the trivialization of $w_2$ correctly.
Summarizing, the computation of the moment $Z_n$ proceeds as follows.
\begin{enumerate}
    \item Firstly, we ignore the imaginary factor on $I_1\cup\overline{I}_1$ in~\eqref{eq:ptrfermi}, and write $Z_n$ without the imaginary factor as the path integral in the form of~\eqref{eq:naiveZn}.
    \item Then, we introduce the effect of the imaginary factor, which shifts $E_{X}$ by the fermion parity twist line $C_{\chi}$. $Z_n$ is eventually expressed as~\eqref{eq:correctZn}.
\end{enumerate}
In the following subsection, we explicitly compute $Z_n$ following the above procedure, dividing into the cases of even or odd $n$.
Though we will mainly work on the Kitaev chain wave function, the following analysis is safely extended for generic spin TQFT prepared by a $\mathbb{Z}_2$-graded Frobenius algebra (see~\ref{app:ptrfrob}).

\subsection{Even powers}
\begin{figure}[htb]
\centering
\includegraphics{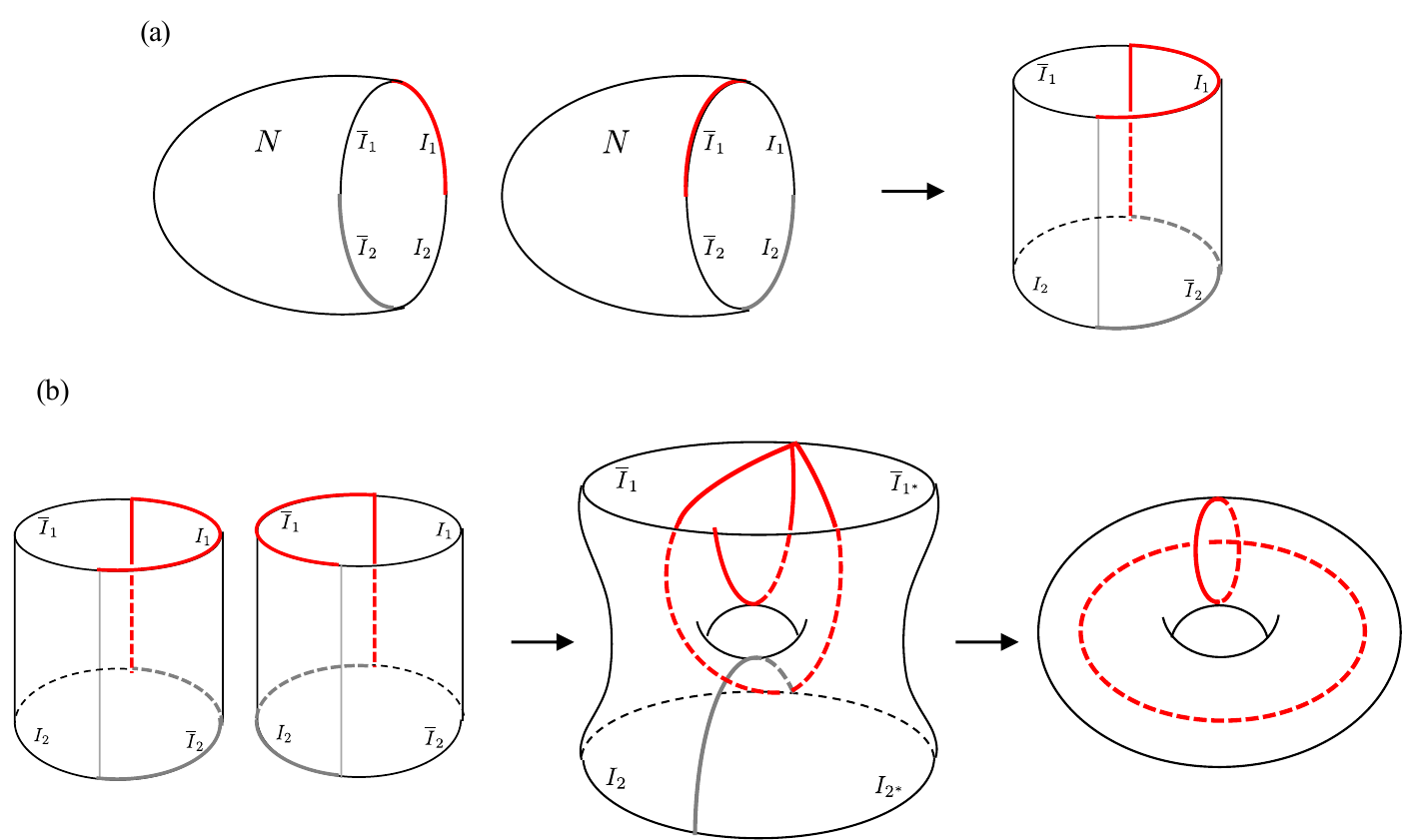}
\caption{Every time we glue surfaces along $I_1$ and $\overline{I}_1$, we introduce the twist by the fermion parity in the resulting surface along the gluing interval.
(a): When we glue two copies of $N$ to take the square of $\rho_I^{\mathcal{T}_1}$, we introduce the fermion parity twist along the red line in the resulting cylinder. The induced spin structure is Ramond. (b): When we glue four copies of $N$ to take the fourth power of $\rho_I^{\mathcal{T}_1}$, we introduce the fermion parity twist along the red line in the resulting punctured torus. If we close the puncture by taking the trace, the induced spin structure on the torus is (R, R).
}
\label{fig:glueRR}
\end{figure}

For even $n$, the resulting surface $X$ becomes an oriented closed surface $\Sigma_g$ with genus $g=n/2-1$, see Fig.~\ref{fig:glueRR} (b) for the case of $n=4$. 
In this case, one can check that $E_{X}$ in~\eqref{eq:naiveZn} correctly provides the trivialization of $w_2$, hence~\eqref{eq:naiveZn} represents a spin TQFT. In the same way as the case of the entanglement negativity, the induced spin structure $\eta_X$ is given by the connected sum of $g$ copies of (NS, NS) torus.

Then, let us consider the effect of the imaginary factor, which makes $Z_n$ different from the case of the entanglement negativity $e^{\mathcal{E}_n}$.
As we discussed above, this factor shifts the spin structure of the resulting surface, by inserting the fermion parity twist in the interval where we glued the copies of $N$, see Fig.~\ref{fig:glueRR} (a). 
For instance, in the case of $n=4$, the twist lines run along two fundamental cycles of the torus, shifting the spin structure from (NS, NS) to (R, R), see Fig.~\ref{fig:glueRR} (b).  
Generally, one can see that $\Sigma_g$ is a connected sum of $(g+1)/2$ copies of tori with (R, R) spin structure, and $(g-1)/2$ copies of tori with (NS, R) spin structure when $n = 0, 4$ mod $8$. 

However, for the case of $n = 2, 6$ mod $8$, the line of the fermion parity twist is no longer closed, which introduces the vortex of the fermion parity at the end of the twist line. This violates the gauge invariance under $\alpha\to\alpha+\delta\lambda$ of the theory~\eqref{eq:pintft}.
Thus after summing over $\alpha\in Z^1(M,\mathbb{Z}_2)$ the partition function becomes zero. Hence, we conclude that $Z_n=0$ when $n = 2, 6$ mod $8$, as shown in~\eqref{eq:momentperiod8}.

Now let us evaluate $Z_n$ for the 
ground state wave function of the Kitaev chain described in Sec.~\ref{subsec:kitaev}, when $n = 0, 4$ mod $8$. The phase of $Z_n$ is given by the Arf invariant of $\Sigma_g$ equipped with the spin structure discussed above. Recalling the Arf invariant on the torus becomes $+1$ for (NS, R) and $-1$ for (R,R), the Arf invariant for $\Sigma_g$ becomes $-1$ for $n=4$ mod $8$, and $+1$ for $n=0$ mod $8$. If we correctly normalize the wave function to match the amplitude, $Z_n$ for $n = 0, 4$ mod $8$ is given by~\eqref{eq:momentperiod8}.

\subsection{Odd powers}
For odd $n$, the resulting surface $X$ is $\Sigma_g$ with genus $g=(n-1)/2$. In this case, $E_X$ in~\eqref{eq:naiveZn} does not give the trivialization of $w_2$. One can check that $\partial E_X$ becomes the set of all 0-simplices of the barycentric subdivision in $X$, except for a pair of 0-simplices
which are denoted as $v_1$ and $v_2$, where the equation $\delta\eta=w_2$ is violated. 

Then, let us incorporate the effect of the fermion parity twist. For odd $n$, the network $C_{\chi}$ of fermion parity twist has two junctions where $n$ twist lines gather at a point. The network becomes not closed at the junctions, and one can check that $v_1, v_2$ are exactly where the two junctions live. Thus, $E_X+C_{\chi}$ in~\eqref{eq:correctZn} correctly gives the spin structure on $X$ after all. The spin structure induced by $E_X+C_{\chi}$ on $X=\Sigma_g$ is determined by the same logic described around \eqref{eq:determineNS}, and summarized as follows.

On one hand, if $n=4m+1$, the spin structure is given by the connected sum of $g/2$ copies of (R, R) tori
and $g/2$ copies of (NS, R) tori.
Especially, for $n=1$, $X$ is a sphere equipped with a spin structure. On the other hand, if $n=4m+3$, the spin structure is given by the connected sum of $(g+1)/2$ copies of (R, R) tori and $(g-1)/2$ copies of (NS, R) tori.

Let us evaluate $Z_n$ for the ground wave function 
of the Kitaev chain described in Sec.~\ref{subsec:kitaev}, when $n = 1,3,5,7$ mod $8$. The phase of $Z_n$ is again given by the Arf invariant of $\Sigma_g$ equipped with the spin structure discussed above. One can see that the Arf invariant for $\Sigma_g$ becomes $-1$ for $n=\pm 3$ mod $8$, and $+1$ for $n=\pm 1$ mod $8$. If we correctly normalize the wave function to match the amplitude, $Z_n$ for $n = 1,3,5,7$ mod $8$ is given by~\eqref{eq:momentperiod8}.

\subsection{Spectrum of partial time-reversal}
Once we have determined the moments for any degree, it is a simple matter to obtain the spectrum of partial time-reversal $\rho_I^{\mathcal{T}_1}$ for Kitaev chain wave function. First, due to the mod 8 periodicity of $Z_n$, the phases of eigenvalues of $\rho_I^{\mathcal{T}_1}$ are all quantized as the eighth root of unity. Since we have $Z_{8n}=4\times (2\sqrt{2})^{-n}$, the spectrum consists of four nonzero values whose absolute values are $1/(2\sqrt{2})$, otherwise zero. By further matching the spectrum with the obtained value of $Z_n$, we see that the four nonzero eigenvalues of $\rho_I^{\mathcal{T}_1}$ are given by
\begin{align}
    \left\{ \frac{1}{2\sqrt{2}} e^{i \pi /4}, \frac{1}{2\sqrt{2}} e^{i \pi /4}, \frac{1}{2\sqrt{2}} e^{-i \pi /4}, \frac{1}{2\sqrt{2}} e^{-i \pi /4} \right\}.
\end{align}

\section{Partial reflection}
\label{sec:pr}
In this section, we discuss another 
quantized non-local order parameter for (1+1)d SPT phases protected by spatial reflection symmetry, proposed in~\cite{Shapourian-Shiozaki-Ryu2017a}.
The proposed quantity is associated with operating the reflection partially on the state. Concretely, let us consider the ground state of the SPT phase $\ket\psi$ protected by the spatial reflection symmetry $\mathcal{R}$, where we have $\mathcal{R}\ket\psi=\ket\psi$.
Then, the ``partial reflection'' is defined as
\begin{align}
    \bra{\psi}\mathcal{R}_{\mathrm{part}}\ket{\psi},
    \label{eq:pr}
\end{align}
where $\mathcal{R}_{\mathrm{part}}$ denotes an operator which reflects a segment in the lattice system. As well as the partial time-reversal in Sec.~\ref{sec:pt}, the partial reflection diagnoses the $\mathbb{Z}_8$ classification of 
(1+1)d topological superconductors 
protected by reflection symmetry $\mathcal{R}$ 
satisfying $\mathcal{R}^2=(-1)^F$.

In the following, we compute the partial reflection~\eqref{eq:pr} on the state prepared by the 
pin$^{-}$ TQFT~\eqref{eq:in},~\eqref{eq:out}, and show that~\eqref{eq:pr} is identical to the partition function of the pin$^{-}$ TQFT,
\begin{align}
    \bra{\psi}\mathcal{R}_{\mathrm{part}}\ket{\psi}=Z[X,\eta],
\end{align}
where $X$ is a closed pin$^-$ surface which generates the pin$^-$ bordism group $\Omega_2^{\mathrm{pin}^-}(\mathrm{pt})=\mathbb{Z}_8$.

\subsection{Evaluation of partial reflection}
We prepare the SPT state $\ket{\mathrm{in}}$ in the form of~\eqref{eq:in}, via the path integral on an open surface $M$. Let the partial reflection $\mathcal{R}_{\mathrm{part}}$ 
act on the interval $I$ in $\partial M$, expressed in Fig.~\ref{fig:MMN}. The action of $\mathcal{R}_{\mathrm{part}}$ on fermion operators 
is then given by
\begin{equation}
\begin{aligned}
& \mathcal{R}_{\mathrm{part}} c_j \mathcal{R}^{-1}_{\mathrm{part}} =
\begin{cases}
i c_{-j} \quad e_j\in I\\
c_{j} \quad \mathrm{otherwise},
\end{cases}\\
& \mathcal{R}_{\mathrm{part}} c_j^{\dagger} \mathcal{R}^{-1}_{\mathrm{part}} =
\begin{cases}
- i c_{-j}^{\dagger} \quad e_j\in I\\
c_{j}^{\dagger} \quad \mathrm{otherwise}.
\end{cases}
\end{aligned}
\end{equation}
Then, the order parameter is given by
\begin{equation}
\begin{aligned}
\braket{ \mathrm{out} | \mathcal{R}_{\mathrm{part}} | \mathrm{in} }
= & \sum_{\alpha \in Z^1(M, \mathbb{Z}_2)} \sum_{\alpha' \in Z^1(\overline{M}, \mathbb{Z}_2)} \widetilde{Z}_{M}[\alpha] \widetilde{Z}_{\overline{M}}[\alpha'] \sigma(M, \alpha; \mathrm{ord}(-n, \cdots, n))\\
& \times \sigma(\overline{M}, \alpha'; \mathrm{ord}(\overline{n}, \cdots, -\overline{n})) (-1)^{\int_{E_M} \alpha} (-1)^{\int_{E_{\overline{M}}} \alpha'}\\
& \times \prod_{1 \leq j \leq l} (-i)^{\alpha(e_j)} (-i)^{\alpha(e_{-j})}
\times \mathrm{sgn}[\alpha(e_{-l}) \cdots \alpha(e_{l})]\\
& \times \prod_{1\leq |j| \leq l} \delta_{\alpha(e_j) \alpha'(e_{-\overline{j}})} \prod_{l+1 \leq |j| \leq n}\delta_{\alpha(e_j) \alpha'(e_{\overline{j}})},
\end{aligned}
\end{equation}
where $\mathrm{sgn}[\alpha_1 \dots \alpha_n]$ is a fermionic sign associated with reordering $\epsilon_1\dots\epsilon_n$ into $\epsilon_n\dots\epsilon_1$, where $\epsilon_i$ carries fermion parity $\alpha_i$.
The expression looks like path integral on the manifold $X$ given by gluing $M$ and $\overline{M}$ via the partial reflection, i.e., identifying the interval $I$ with $\overline{I}$ by the orientation reversing map and the complement by the orientation preserving map. Here, the induced map $M\sqcup \overline{M}\to X$ is restricted to $M$ as $M\to \widetilde{M}$, where $\widetilde{M}$ is given by identifying two 0-simplices contained in $\partial I$ on $\partial M$, since these two 0-simplices are identified in $X$.
Similarly, we also have the restriction $\overline{M}\to \overline{\widetilde{M}}$.

Since the product of the shadow theories on $M$ and $\overline{M}$ defines a shadow theory on $X$, one can rewrite the above expression as 
\begin{equation}
\begin{aligned}
\braket{ \mathrm{out} | \mathcal{R}_{\mathrm{part}} | \mathrm{in} }
= & \sum_{\alpha \in Z^1(X, \mathbb{Z}_2)} \widetilde{Z}_{X}[\alpha] \sigma(\widetilde{M}, \alpha|_{\widetilde{M}}; \mathrm{ord}(-n, \cdots, n)) \sigma(\overline{\widetilde{M}}, \alpha|_{\overline{\widetilde{M}}}; \mathrm{ord}(\overline{n}, \cdots, -\overline{n}))\\
& \times (-1)^{\int_{E_{\widetilde{M}}} \alpha|_{\widetilde{M}}} (-1)^{\int_{E_{\overline{\widetilde{M}}}} \alpha|_{\overline{\widetilde{M}}}}
\prod_{e\in I} (-i)^{\alpha(e)}
\times \mathrm{sgn}[\alpha(e_{-l}) \cdots \alpha(e_{l})],
\label{eq: many-body top R}
\end{aligned}
\end{equation}
where $\alpha|_{\widetilde{M}}$ (resp. $\alpha|_{\overline{\widetilde{M}}}$) represents the restriction of $\alpha \in Z^1(X, \mathbb{Z}_2)$ to $\widetilde{M}$ (resp. $\overline{\widetilde{M}}$), and $E_{\widetilde{M}}$ is an image of $E_{M}$ under $M\to \widetilde{M}$.
We can show that the quantity~\eqref{eq: many-body top R} gives the partition function of the pin$^-$ invertible TQFT defined on $X$. This is done by comparing~\eqref{eq: many-body top R} with the partition function of the pin$^-$ TQFT given by
\begin{equation}
\begin{aligned}
Z[X, \eta] = & \sum_{\alpha \in Z^1(X, \mathbb{Z}_2)} \widetilde{Z}_{X}[\alpha] \sigma(\widetilde{M}, \alpha|_{\widetilde{M}}; \mathrm{ord}(-n, \cdots, n)) \sigma(\overline{\widetilde{M}}, \alpha|_{\overline{\widetilde{M}}}; \mathrm{ord}(\overline{n}, \cdots, -\overline{n}))\\
& \times (-1)^{\int_{E_X} \alpha}
\prod_{l+1 \leq j \leq n}^{\mathrm{odd}}(-1)^{\alpha(e_j)}
\prod_{l+1 \leq j \leq n}^{\mathrm{even}}(-1)^{\alpha(e_{-j})}\\
&\times \prod_{e\in I}(-i)^{\alpha(e)}\times \mathrm{sgn}[\alpha(e_{-l}) \cdots \alpha(e_{l})].
\label{eq: Z[X]2}
\end{aligned}
\end{equation}
Comparing the equation~\eqref{eq: many-body top R} with~\eqref{eq: Z[X]2}, one finds that these expressions are identified if
\begin{equation}
E_{\widetilde{M}}+E_{\overline{\widetilde{M}}}+\sum_{l+1 \leq j \leq n}^{\mathrm{odd}} e_j
+\sum_{l+1 \leq j \leq n}^{\mathrm{even}} e_{-j}=E_X,
\label{eq:refdual}
\end{equation}
where $\partial E_X$ is the dual of $w_2+w_1^2$ on $X$, specified in Sec.~\ref{subsec:pintft}. Here, we note that the boundary contribution of $E_{\widetilde{M}}$, $E_{\overline{\widetilde{M}}}$ cancel out, once we postulate the reflection symmetry of the state $\mathcal{R}\ket{\mathrm{in}}=\ket{\mathrm{in}}$, making $\partial E_M$ reflection symmetric on $\partial M$. Then, we can check that the lhs of~\eqref{eq:refdual} gives the correct trivialization of $w_2+w_1^2$ on $X$. Hence, we have shown that
\begin{align}
   \braket{ \mathrm{out} | \mathcal{R}_{\mathrm{part}} | \mathrm{in} }=Z[X, \eta].
\end{align}
For instance, let us evaluate the quantity in the wave function of the Kitaev chain described in Sec.~\ref{subsec:kitaev}. Using the form of the ABK invariant~\eqref{eq:abk}, the expression becomes
\begin{equation}
\braket{ \mathrm{out} | \mathcal{R}_{\mathrm{part}} | \mathrm{in} } = 2^{-\chi(Y)/2 + \chi(X) / 2} \mathrm{ABK}[X, \eta],
\end{equation}
where $Y$ is a closed surface given by gluing $M$ and $\overline{M}$ along the boundaries.
In particular, if we choose $M$ as a disk, $Y = S^2$ and $X = \mathbb{RP}^2$, which gives 
\begin{equation}
\braket{ \mathrm{out} | \mathcal{R}_{\mathrm{part}} | \mathrm{in} } = \frac{1}{\sqrt{2}} e^{\pm 2\pi i / 8}.
\label{eq:partref}
\end{equation}
This reproduces the results obtained in \cite{Shapourian-Shiozaki-Ryu2017a}. 

\section{Conclusions}
\label{sec:conclusions}
In conclusion, we explicitly computed entanglement measures and quantized non-local order parameters for fermionic SPT phases in the framework of spin/pin TQFT. 
We clarified the properties of 
order parameters defined via partial operations, as topological invariants diagnosing the $\mathbb{Z}_8$ classification of (1+1)d fermionic SPT phases in class BDI and D$+$R$_-$. 
Moreover, we demonstrated that these 
order parameters have universal amplitudes, indicating a topological origin such as the quantum dimension of the boundary Majorana modes.
Furthermore, we revealed that the moments of partial time-reversal have the mod 8 periodicity, which leads to the eight-fold quantization of the negativity spectrum.

There are several avenues to pursue for future research.
First, a natural extension of the present paper is to explore the formulation in higher dimensions.
In this case, we should be able to prepare a wave function in the form of tensor network state, from a path integral of spin/pin TQFT in generic dimension.
In higher dimensions, it is suggested \cite{Shiozaki-Shapourian-Ryu2017} that fermionic SPT phases with point group symmetries can be detected by partial point group operations. It is interesting to formulate the partial point group operation and its relationship to path integral of lattice TQFT. 
It also remains open for future works to examine entanglement properties of spin/pin TQFT in higher dimensions.

Furthermore, it is worth investigating entanglement measures studied in the present paper for conformal field theory (CFT) coupled with a spin structure. Since we can obtain a fermionic CFT by coupling a bosonic CFT with a spin/pin TQFT, we believe that our formulation of entanglement measures is useful in studying entanglement properties of fermionic CFT.
Especially, it was demonstrated in \cite{Shapourian-Ruggiero-Ryu-Calabrese2019} that a critical Majorana chain with $c=1/2$ shows the six-fold quantization of the negativity spectrum, which resembles the eight-fold quantization in spin TQFT discovered in the present paper. It is conceivable that the six-fold quantization for eigenvalues of $\rho_I^{\mathcal{T}_1}$ is a universal nature of spin CFT not limited to a Majorana chain, since the moment of partial time reversal is associated with a three point function of twist fields of CFT. The derivation of the six-fold quantization for the case of fermionic CFT is left for a future work. 

%

\section*{Acknowledgments} 
We thank Hassan Shapourian
for useful discussions.
RK acknowledges the hospitality of Harvard CMSA.
KI is supported by Program of Excellence in Photon Science (XPS). 
RK is supported by Japan Society for the Promotion of Science (JSPS) through Grant No.~19J20801. 
SR is supported by the National Science
Foundation under award number DMR-1455296, and by
a Simons Investigator Grant from the Simons Foundation.

\appendix
\section{Partial time-reversal: Cases with general Frobenius algebra}
\label{app:ptrfrob}
In this appendix, we formulate partial time-reversal for spin/pin TQFT wave function prepared by a $\mathbb{Z}_2$ graded Frobenius algebra $A$. In a similar way to Sec.~\ref{subsec:ptrcomp}, we prepare the Hilbert space on $S^1=\partial M$, and let $e_{-n},\dots,e_n$ as boundary 1-simplices. Then, the state is expressed like~\eqref{eq:in},~\eqref{eq:out} as
\begin{align}
& \ket{\mathrm{in}} = \sum_{\alpha \in Z^1({M}, \mathbb{Z}_2)}\sum_{A^{\widehat{\otimes}2n}} \widetilde{Z}_{M}[\alpha] \sigma(M, \alpha; \mathrm{ord}(-n,\cdots,n)) (-1)^{\int_M \eta \cup \alpha} \ket{\epsilon_{-n}\dots \epsilon_{-1}\epsilon_{1}\dots \epsilon_{n}},
\label{eq:infrob}\\
& \bra{\mathrm{out}} = \sum_{\alpha \in Z^1({\overline{M}}, \mathbb{Z}_2)}\sum_{A^{\widehat{\otimes}2n}} \widetilde{Z}_{\overline{M}}[\alpha] \sigma(\overline{M}, \alpha; \mathrm{ord}(\overline{n}, \cdots, -\overline{n})) (-1)^{\int_{\overline{M}} \eta \cup \alpha} \bra{\epsilon_{\overline{n}}\dots \epsilon_{\overline{1}}\epsilon_{-\overline{1}}\dots \epsilon_{-\overline{n}}},
\label{eq:outfrob}
\end{align}
where the sum for $A^{\widehat{\otimes}2n}$ runs over the elements $\ket{\epsilon_{-n}\dots \epsilon_{-1}\epsilon_{1}\dots \epsilon_{n}}$ with a fixed $\mathbb{Z}_2$ grading specified by $\alpha$ on the boundary. To formulate the partial time-reversal, we need to give a definition of the inner product and the trace. We postulate that
\begin{align}
  \langle \epsilon_i|\epsilon_j \rangle =\mathrm{tr}(\ket{\epsilon_{i}}\bra{\epsilon_{j}})=g^{ij},
\end{align}
where $g^{ij}$ is the weight on 1-simplices in the state sum of the shadow theory $\widetilde{Z}$. When the above inner product is a sesquilinear positive definite form, and further $A$ is a commutative Frobenius algebra equipped with a structure of a $\ast$-algebra $C^i_{jk}=C^{i*}_{kj}$, then
the theory is guaranteed to be unitary~\cite{Kapustin-Turzillo-You2017}. In the following, we assume that $A$ satisfies these properties.
Then, replicating the logic of Sec.~\ref{subsec:ptrcomp}, the reduced density matrix for the interval $I=\bigcup_{1\le |j|\le l}e_j$ is given in the form of~\eqref{eq:reducedN},
\begin{equation}
\begin{aligned}
\rho_I = & \sum_{\alpha \in Z^1(N, \mathbb{Z}_2)}\sum_{A^{\widehat{\otimes}4l}} \widetilde{Z}_N[\alpha] \sigma(N, \alpha; \mathrm{ord}(-l, \cdots, l, \overline{l}, \cdots, -\overline{l})) (-1)^{\int_{E_N} \alpha}\\
& \times \ket{\epsilon_{-l}\dots \epsilon_{l}}\bra{\epsilon_{\overline{l}}\dots \epsilon_{-\overline{l}}}.
\end{aligned}
\end{equation}
Imitating the case of $A=Cl(1)$ expressed in~\eqref{eq:ptrfermi}, we define the partial time-reversal for $\rho_I$ as
\begin{equation}
\begin{aligned}
\rho_I^{\mathcal{T}_1} = & \sum_{\alpha \in Z^1(N, \mathbb{Z}_2)} \widetilde{Z}_N[\alpha] \sigma(N, \alpha; \mathrm{ord}(-l, \cdots, -1, \overline{1}, \cdots, \overline{l},l,\cdots, 1,-\overline{1}, \cdots, -\overline{l}))\\
&\times(-1)^{\int_{E_N} \alpha}
 \prod_{e\in I_1\cup\overline{I}_1} (-i)^{\alpha(e)} 
 \times \ket{\epsilon_{-l}\dots \epsilon_{-1}\epsilon_{\overline{1}}\dots \epsilon_{\overline{l}}}\bra{\epsilon_{l}\dots \epsilon_{1}\epsilon_{-\overline{1}}\dots \epsilon_{-\overline{l}}}.
\end{aligned}
\end{equation}
Based on the above definition of partial time-reversal, we can compute the entanglement negativity and moment of partial time-reversal for generic wave function of spin TQFT in the same fashion as Sec.~\ref{sec:nega},~\ref{sec:moment}. 
It is straightforward to see that the results presented in Sec.~\ref{sec:nega},~\ref{sec:moment} are also true for generic spin TQFT on lattice. The results in Sec.~\ref{sec:pt},~\ref{sec:pr} are also extended to pin$^-$ TQFT on lattice constructed from $A$.

\section*{References}

\bibliography{ref}

\end{document}